\documentclass[twocolumn]{aastex62}

\usepackage{amsmath}
\usepackage{comment}
\usepackage[inline]{enumitem}

\usepackage{natbib}
\usepackage{array,multirow}

\newcommand{\obliq}{\epsilon}
\newcommand{\axaz}{\zeta}
\newcommand{\HD}{HD 149026~b}
\newcommand{\WASP}{WASP-12~b}
\newcommand{\COROT}{CoRoT-2~b}

\received{2019 March 18}
\accepted{2019 June 18}
\submitjournal{AJ}

\shorttitle{Phase Curves of Oblique Planets}
\shortauthors{Adams, Millholland, \& Laughlin}

\begin{document}

\title{Signatures of Obliquity in Thermal Phase Curves of Hot Jupiters}

\author[0000-0002-7139-3695]{Arthur D. Adams}
\affiliation{Department of Astronomy, Yale University, New Haven, CT 06520}
\author[0000-0003-3130-2282]{Sarah Millholland}
\affiliation{Department of Astronomy, Yale University, New Haven, CT 06520}
\author[0000-0002-3253-2621]{Gregory P. Laughlin}
\affiliation{Department of Astronomy, Yale University, New Haven, CT 06520}

\begin{abstract}
Recent work suggests that many short-period extrasolar planets may have spin obliquities that are significantly tilted with respect to their orbital planes. These large obliquities are a natural outcome of ``secular spin-orbit resonance'', a configuration in which the planetary spin precession frequency matches the frequency of orbit nodal regression, or a Fourier component thereof. While exoplanet spin obliquities have not yet been measured directly, they may be detectable indirectly through their signatures in various observations, such as photometric measurements across the full phase of a planet's orbit. In this work, we employ a thermal radiative model to explore how large polar tilts affect full-phase light curves, and we discuss the range of unique signatures that are expected to result. We show that the well-studied short-period planets \HD{}, \WASP{}, and \COROT{} all exhibit phase curve features that may arise from being in high-obliquity states. We also constrain the parameters and assess the detectability of hypothetical perturbing planets that could maintain the planets in these states. Among the three planets considered, \COROT{} has the tightest constraints on its proposed obliquity ($45.8^{\circ} \pm 1.4^{\circ}$) and axial orientation. For \HD{}, we find no significant evidence for a non-zero obliquity, and the phase curve of \WASP{} is too complicated by strong tidal distortions for a conclusive assessment.   
\end{abstract}


\section{Introduction}\label{sec:introduction} 

Analysis of the full-phase photometry of transiting extrasolar planets has generated a number of insights. By tracking the emission and reflection of light from planets as they trace through their orbits, one may derive important clues regarding the planets' atmospheric compositions, surface flow patterns, atmospheric thermal responses, cloud coverage, day-to-night heat redistribution efficiency, and more. High signal-to-noise detections of full-phase photometry are now routinely performed in both the optical and in the near-infrared, and the technique has been employed for both detection and characterization \citep[see, e.g.][]{heng15,shp17,dem17}.

The climate properties probed by full-phase light curves are highly sensitive to the planet's orbital and spin geometry. For instance, many short-period ($P\lesssim 5\,{\rm d}$) planets are subject to strong tidal evolution that has produced complete or near-complete orbital circularization \citep{Rod2010}. Moreover, because time scales for planetary tidal spin evolution are generally shorter than for orbital evolution, it is usually assumed that short-period planets have zero obliquity, and are spinning synchronously for orbits with $e=0$ \citep[see e.g.][]{Gladman1996} or pseudo-synchronously for orbits with finite eccentricity \citep[see e.g.][]{hut81,iva07}.

Time-resolved photometry of assumed-synchronous planets on circular orbits suggests that the peak infrared emission from the planet usually lies eastward of the sub-stellar point \citep[e.g.][]{knu07b,knu09a,cow11a,knu12,Cowan2012,zel14,sch17,zha18}. By contrast, phase curves in the optical tend to suggest that peak reflectivity occurs westward of the sub-stellar point and closer to the morning terminator \citep{shp17}. These observations are generally interpreted to imply eastward circumplanetary flow and cloud decks that burn off when advected into the direct beam of instellation.

This composite picture can be tested when the phase curves of eccentric orbits are tracked \citep[e.g.][]{lau09,Adams2019a}. A planet with a non-circular orbit cannot be fully tidally de-spun; the surface flows are thus dynamically responsive. Bulk properties of the atmosphere such as the radiative response timescale can be directly inferred \citep[e.g.][]{Cowan2012,dew16,ada18b}. 

Just as the eccentricity produces observable signatures in the full-phase photometry, so too will the presence of a significant obliquity. It is thus of interest to examine possible obliquity-induced signatures in the context of the data sets that are currently available.

Several authors have developed a mathematical formalism to predict optical observational effects of the relationship between a planet's spin, orbit, and viewing geometry. \citet{Kawahara2010} constructed a geometric framework for mapping planets (including oblique ones) in reflected/scattered light. This framework was later extended to account for effects such as cloud cover \citep{Kawahara2011} and generally inclined orbits \citep{Fujii2012}. Signatures of obliquity have been predicted in Fourier analyses of photometry for directly-imaged planets \citep{Kawahara2016}. \citet{Schwartz2016} demonstrated the ability to infer spin axis orientation for general albedo maps, and outlined a feasible minimum observing baseline for making robust inferences of the spin axis orientation. \citet{Farr2018} introduced \texttt{exocartographer}\footnote{\url{https://github.com/bfarr/exocartographer}}, a software package that generates reflected light photometry for an arbitrary albedo map and spin geometry and fits a variety of time-sampled data. Another software package, \texttt{STARRY}\footnote{\url{https://rodluger.github.io/starry/}} \citep{lug19}, provides computationally efficient determinations of full-phase light curves, occultation signatures, and transit signals by using a global planetary surface pattern expressed as a sum of spherical harmonics.

Here we consider a similar geometric framework in the near-infrared, where the thermal emission from the planet --- rather than reflected light --- should be the primary source of the observed flux. \citet{Cowan2012} considered the effects of eccentricity and obliquity for the phase photometry of an Earth-like planet and demonstrated that one can infer the characteristic thermal time scales. They also concluded that a combination of optical and infrared observations would be necessary to accurately measure bulk atmospheric conditions. Most recently, \citet{Ohno2019a,Ohno2019b} developed a comprehensive shallow water model for describing the atmospheric dynamics and resulting thermal phase variations of planets with arbitrary spin period, spin obliquity, and orbital eccentricity. In short, because these three properties all determine the instantaneous sub-stellar location, the resulting thermal phase variations will be shaped by their combined influence.

While one can model thermal variations for any choice of orbit and spin orientation, we will first restrict ourselves to scenarios that are physically feasible for systems with close-in giant planets. In addition to representing a dynamically plausible configuration, accurate phase curves for short-period giant planets already exist. Moreover, if we infer a particular spin-orbit geometry for a planet from its observed photometry, we may place constraints on the system architecture and make potentially observable dynamical predictions.

In this work, we focus on three short-period giant planets: \HD{}, \WASP{}, and \COROT{}. These all have full-phase thermal light curves obtained with \textit{Spitzer}, and they have features \citep[discussed in detail for \HD{} and \WASP{} in][]{ada18b} that suggest they are potential high obliquity candidates. \HD{}, \WASP{}, and \COROT{} do not currently have any known planetary companions. They would therefore require special dynamical states to have anything other than circular orbits, spin-orbit synchronization, and zero obliquities. 

One such dynamical state that may maintain a large obliquity is a ``Cassini state'', an equilibrium configuration where the planet's spin vector stays fixed in the reference frame of its precessing orbit \citep{Colombo1966,Peale1969,Ward1975a}. 
As an instance of a Cassini state, a secular spin-orbit resonance involves an average commensurability between the frequency of the planet's spin axis precession and the frequency of its orbit nodal recession (or a Fourier component thereof). The orbital recession may be provided by a number of sources, such as another planet in the system, the stellar quadrupolar gravitational potential \citep{Fabrycky2007, MillhollandLaughlin2019}, or, early on in the system's lifetime, the protoplanetary disk \citep{MillhollandBatygin2019}. The result is a stable state in which a planet may maintain a non-synchronous spin and non-zero obliquity, even in the presence of strong tides. 

Cassini states have been invoked to explain, for example, the co-precession of the lunar spin vector and the lunar orbit normal \citep{DeLaunay1860}, the obliquities of Saturn \citep{Ward2004,Hamilton2004} and Jupiter \citep{Ward2006}, and the spin precession state of Mercury \citep{Peale2006}. They have also been considered as a mechanism to inflate the radii of hot Jupiters \citep{Winn2005}, albeit with disputed feasibility \citep{Levrard2007,Fabrycky2007}. Recently, \cite{MillhollandLaughlin2019} showed that high-obliquity Cassini states might be common for planets in short-period, compact systems. In this paper, we examine whether \HD{}, \WASP{}, and \COROT{} may have large obliquities due to their participation in secular spin-orbit resonances, and we examine the potential signatures of such states in their thermal full-phase light curves.

This paper is organized as follows. In \S \ref{sec: Three Unusual Planets}, we introduce three unusual planets -- \HD{}, \WASP{}, and \COROT{} -- and discuss why they are viable candidates for obliquity investigations. In \S \ref{sec:photometry} we review these planets' near-infrared photometric observations, including the analyses of their light curve morphologies. \S \ref{sec:model} describes the thermal model, which accounts for a general rotation rate and spin axis orientation. This framework is then employed in \S \ref{sec:results} to re-analyze the \textit{Spitzer} phase photometry. In order to build a physical framework where we could plausibly observe a single-transiting, close-in, oblique planet, we make a feasibility assessment in \S \ref{sec:REBOUND} wherein a compact, nearly co-planar multi-planet system evolves on short timescales to states of high mutual inclination, with only one planet capable of transiting. Finally, \S \ref{sec:Cassini state} provides a dynamical analysis of the three case study systems and explores the possibility that a Cassini state may exist between the known planets and an as-yet undiscovered, non-transiting companion.

\section{Three Unusual Planets}\label{sec: Three Unusual Planets}

\begin{deluxetable*}{cccc}
\centering
\tabletypesize{\footnotesize}
\tablecaption{Planetary System Properties}
\tablewidth{0pt}
\tablehead{\colhead{} & \colhead{\HD{}} & \colhead{\WASP{}} & \colhead{\COROT{}}}

\startdata
$P$ (days) & $2.8758911 \pm 2.5\times10^{-6}$ & $1.09142119 \pm 2.1\times10^{-7}$ & $1.7429935 \pm 1.0\times10^{-6}$ \\
$e$ & $0$ & $0.0447\pm0.0043$ & $0.0143^{+0.0077}_{-0.0076}$ \\
$\varpi\left(^\circ\right)$ & N/A & $272.7^{+2.4}_{-1.3}$ & $102^{+17}_{-5}$ \\
$M_\mathrm{P}$ $\left(M_\mathrm{J}\right)$ & $0.368^{+0.013}_{-0.014}$ & $1.43 \pm 0.14$ & $3.47 \pm 0.22$ \\
$R_\mathrm{P}$ $\left(R_\mathrm{J}\right)$ & $0.813^{+0.027}_{-0.025}$ & $1.825 \pm 0.094$ & $1.466^{+0.042}_{-0.044}$ \\
$M_\star$ $\left(M_\odot\right)$ & $1.345 \pm 0.020$ & $1.280 \pm 0.05$ & $0.96 \pm 0.08$ \\
$R_\star$ $\left(R_\odot\right)$ & $1.541^{+0.046}_{-0.042}$ & $1.630 \pm 0.08$ & $0.906^{+0.026}_{-0.027}$ \\
$T_{\textrm{eff}}$ (K) & $6160 \pm 50$ & $6300^{+200}_{-100}$ & $5625 \pm 120$ \\
Ref. & (1) & (2)--(4) & (5) \\
\enddata

\tablerefs{(1) \citet{Carter2009}; (2) \citet{tur16} ($P$, $e$); (3) \citet{knu14} ($\varpi$); (4) \citet{sou12}; (5) \citet{Gillon2010}.}

\label{table:planet_properties}
\end{deluxetable*}

\HD{} orbits its subgiant host in 2.88 days; the planet is slightly more massive (0.38 M$_\mathrm{J}$) but smaller in size (0.74 R$_\mathrm{J}$) than Saturn \citep{Stassun2017}. Its formation is still a matter of ongoing study, since the bulk density is quite high compared with other short-period giant planets. \citet{Sato2005}, who first measured the planet's radius during transit, proposed that a high core mass of $\sim\!67$ M$_\oplus$ could explain the measurements. Subsequent assessments have inferred similarly high core masses in the range of $\sim\!50$--110 M$_\oplus$ \citep{Fortney2006,Ikoma2006,Broeg2007,Burrows2007} using combinations of atmospheric and interior modeling. \citet{Ikoma2006} proposed that \HD{}'s high metallicity and modest H/He envelope might either be explained via planetesimal capture and a limited gas supply, if its current state was obtained prior to disk dissipation, or a combination of envelope photoevaporation, Roche lobe overflow, and major collisions after disk dissipation. \citet{zha18} find that the best-fit phase offsets from the \textit{Spitzer} 3.6 $\mu$m and 4.5 $\mu$m full phase photometry are both significantly different from zero and in disagreement with each other. While \citet{zha18} suggested uncharacterized instrumental systematics as a potential cause of the disparity, it is also worth investigating whether unusual global atmospheric dynamics may be at play.

\WASP{} is another short-period ($1.09$ days) giant planet, which is tidally distorted due to its density and proximity to its host star \citep{lis10,lai10}. There is spectroscopic evidence that the planet is overflowing its Roche lobe \citep{Fossati2010,Haswell2012,Fossati2013,Jackson2017}, and hydrodynamic simulations \citep{Debrecht2018} suggest that it is undergoing significant atmospheric mass loss. \citet{VonEssen2019} have recently analyzed time variability in the measured optical eclipse depths; with neither cloud albedo nor high temperatures tenable to explain the variability without invoking extreme physical magnitudes, they point cautiously to additional occultation from the atmospheric mass loss as a possible mechanism. Extensive transit observations have also been made (detailed further in \S \ref{sec:photometry}), whose rapidly advancing ephemerides suggest either apsidal precession of an eccentric orbit or orbital decay \citep{mac2016, Patra2017}. Recent observations \citep{Maciejewski2018} and theoretical investigations \citep{Bailey2019} are in favor of the orbital in-spiral scenario. \cite{MillhollandLaughlin2018} proposed that this rapid orbital decay may be due to tidal dissipation in the planet that is strongly enhanced by a high obliquity state.

\COROT{} is a $3.3 \ M_{\mathrm{J}}$ hot Jupiter with a 1.74-day orbital period and $1.47 \ R_{\mathrm{J}}$ radius that has consistently been measured as anomalously inflated\footnote{It is important to note that the host star is quite active \citep{Lanza2009,Huber2009}, and this activity can bias measurements of the planetary radius from transit observations \citep{Czesla2009,Huber2010,Bruno2016}.} \citep{Alonso2008,Gillon2010,sou11}.  Previous proposed explanations include that the system is very young ($\sim\!30$--40 Myr) and the planet has not yet fully contracted gravitationally, or it is a bit older ($\sim\!130$--500 Myr) and the radius inflation is a transient response from a recent collision \citep{Guillot2011}. Alternatively, the extreme radius inflation may be the result of obliquity tides, a possibility we discuss later in \S \ref{sec:discussion}. In addition, \citet{Dang2018} reported \textit{Spitzer} thermal phase observations that robustly indicate that the day-side hotspot is westward of the sub-stellar point, in contrast to the eastward offset that planets typically exhibit. Magnetohydrodynamic effects may be one mechanism for generating an unusual westward offset, but recently \cite{Hindle2019} inferred that such effects are very unlikely to be a viable mechanism for the specific westward offsets seen at 4.5 $\mu$m in \COROT{}. \citet{Dang2018} considered that it may arise from non-synchronous rotation, which, as discussed in \S \ref{sec:Cassini state}, would be a consequence of a high-obliquity state.

\section{Photometry}\label{sec:photometry}
\HD{} has been observed in transit in Str\"{o}mgren $b$ and $y$ \citep{Sato2005,Winn2008}, $g$ and $r$ \citep{Char2006}, NICMOS (1.1--2.0 $\mu$m) on the Hubble Space Telescope \citep{Carter2009}, and the 8.0 $\mu$m channel of the Infrared Array Camera (IRAC) on the \textit{Spitzer} Space Telescope \citep{Nutzman2009}. Secondary eclipses have also been observed in each of the 4 IRAC channels (3.6--8.0 $\mu$m) and the Infrared Spectrograph (IRS) at 16 $\mu$m \citep{ste12}. \citet{knu09b} presented the first phase photometry, which spanned just over half the orbit in 8.0 $\mu$m. Most recently \citet{zha18} published two full-phase observations in the Warm \textit{Spitzer} bands (3.6 and 4.5 $\mu$m). We draw attention to \S 4.1 of \citet{zha18}, where the authors point out that inconsistencies between the bands warrant a fair degree of skepticism. In particular, the positive phase offset, or late minimum, of the 3.6 $\mu$m time series, is difficult to explain with modeling that assumes spin-orbit synchronization.

\WASP{} was first discovered in transit via the SuperWASP camera \citep{heb09}, and subsequent transits have been observed in the $V$ band \citep{cha12}, $R$ \citep{mac11}, $J$, $H$, and $K_s$ \citep{cro11}, and at 3.6--8.0 $\mu$m from \textit{Spitzer} \citep{cam11,cow12,ste14}. \citet{man13} also provided transit spectroscopy from WFC3 (1.1--1.7 $\mu$m) on the Hubble Space Telescope. Our work focuses on the full-orbit phase curves available from the warm \textit{Spitzer} (3.6 and 4.5 $\mu$m) channels, originally published in \citet{cow12}.

\COROT{} has been observed in transit \citep{Alonso2008} and eclipse via photometry in the \textit{Spitzer} IRAC 3.6 \citep{Deming2011}, 4.5, and 8.0 $\mu$m \citep{Gillon2010} channels. Transit spectra have also been measured \citep{Bouchy2008,Czesla2012,bal15}. Refinement of the stellar parameters for CoRoT-2 led to revised transit depths \citep{sou11,sou12,bal15}. Even with the revised radius, \COROT{} is estimated to have a density comparable with Jupiter's, hinting at a possible radius inflation that might be due to uncharacterized tidal processes. Recently, \citet{Dang2018} published a full-phase light curve at 4.5 $\mu$m, which showed an unusual and substantial $23^{\circ}$ westward phase offset. The authors proposed possible astrophysical sources of the offset including westward winds, magnetic effects, or partial cloud cover. Inhomogeneous cloud cover has been studied as a possible explanation of the westward offsets seen in optical light curves \citep{shp17}, and optical follow-up for \COROT{} may support this hypothesis \citep{Barstow2018}. 
Here we offer the alternative hypothesis of a high obliquity state. We show that this could not only produce the observed westward offset, but it would also provide new information about the formation and dynamical evolution mechanisms of close-in giants.

\section{Components of the Oblique Thermal Model}\label{sec:model}
In order to develop a model of thermal phase variations for a planet with non-zero spin obliquity, we build upon the model framework developed for the analysis in \citet{ada18b}. Each planet is divided into a grid with cells of dimension $5^\circ\times 5^\circ$ by latitude and longitude. We start with the known system properties of each modeled planet (Table \ref{table:planet_properties}). There are 6 tunable parameters which jointly govern the resulting thermal emission. The first three --- the albedo $A$, equilibrium radiative timescale $\tau_{\mathrm{rad}}$, and minimum temperature $T_0$ --- most directly control the thermal properties of each cell. They are put into their formal context in \S \ref{sec:model:thermal}. In short, each cell absorbs a fraction $1-A$ of radiation from the host star and re-radiates as a blackbody with a corresponding characteristic timescale. Its brightness temperature is set both by the time-dependent instellation and any non-stellar heating (e.g. tidal heat emanating from the planet interior), the latter of which is captured with the minimum temperature. The remaining three parameters define the components of the rotation vector. Movement of the cells in the model is set entirely by the rotation; the infrared photospheric layer of the planet rotates with some angular frequency $\omega_{\mathrm{rot}}$ ($\equiv 2\pi/P_{\mathrm{rot}}$ for the rotation period $P_{\mathrm{rot}}$) with some orientation of its axis relative to the orbital plane. This may be written in terms of two spherical angles: $\obliq$, which is the angle between the planet's axis and the orbit normal, and $\axaz$, which is the projected angle of the axis in the orbital plane relative to a reference direction. We set this reference direction along the planet-star line during periastron, such that $\axaz=0$ implies the northern hemisphere\footnote{The northern hemisphere is defined according to the axis around which the planet rotates counter-clockwise. Retrograde motion is therefore obtained by setting $\obliq>90^\circ$.} summer solstice occurs during periastron \citep[equivalent to $f_\mathrm{sol}$ in][]{Ohno2019a}. This corresponds to an angle of the axial projection from the line of sight of $\axaz_\mathrm{obs} = \axaz-\nu_\mathrm{tra}$, for true anomaly at transit $\nu_\mathrm{tra} \equiv \pi/2 - \varpi$; we assume $\varpi \rightarrow \pi/2$ for circular orbits so that $\axaz=\axaz_\mathrm{obs}$. The time evolution of grid cell temperatures is convolved with both the viewing geometry from Earth, which is approximated as edge-on, and the relevant instrumental band profiles, to predict full-orbit light curves for a given set of parameter values. Our fitting routine is detailed in \S \ref{sec:results}.

\subsection{Thermal Evolution of the Cells}\label{sec:model:thermal}
The planetary model is initialized at apastron with a uniform surface temperature $T_0$; for planets on circular orbits, we set an arbitrary argument of periastron $\varpi=\pi/2$ such that apastron occurs during secondary eclipse. To calculate the incoming stellar radiation over the orbit, we start with the star-planet separation
\begin{equation}\label{eq:orbit_distance}
	r\!\left(t\right) = a \left( \frac{1-e^2}{1+e\cos\nu} \right)
\end{equation}
where $a$ is the orbital semi-major axis, $e$ the orbital eccentricity, and $\nu = \nu\!\left(t\right)$ the true anomaly. To solve for the true anomaly from the time in orbit we first calculate the mean anomaly $M$, which is directly proportional to time: $M\!\left(t\right) = \omega_{\mathrm{rot}}\left(t - t_{\mathrm{peri}}\right)$ for rotation rate $\omega_{\mathrm{rot}}$ and periastron passage time $t_{\mathrm{peri}}$. The eccentric anomaly $E=E\!\left(t\right)$ is then given by Kepler's equation
\begin{equation}\label{eq:Kepler}
	M\!\left(t\right) = E - e \sin E.
\end{equation}
There is a direct relation between regular time intervals and regular intervals in mean anomaly, but not for the eccentric or true anomalies; we must calculate the latter numerically. The sine and cosine of the true anomaly, $\nu$ are given by
\begin{align}\label{eq:true_anomaly}
\begin{split}
\cos\nu &= \frac{\cos E - e}{1-e\cos E}\\
\sin\nu &= \frac{\sqrt{1-e^2}\sin E}{1-e\cos E}.
\end{split}
\end{align}

At each time $t$ in the orbit and longitude/latitude $\left(\phi, \theta\right)$ on the planetary surface, the equilibrium temperature is calculated via
\begin{equation}\label{eq:Teq}
    T_{\textrm{eq}}^4\!\left( \phi, \theta, t  \right) = \left(1 - A\right) \left( \frac{L_\star}{4\pi \sigma r^2} \right) \cos\alpha_\star + T_0^4
\end{equation}
where $A$ is the planetary albedo, $L_\star$ the stellar luminosity, $\sigma$ the Stefan-Boltzmann constant, $r=r\!\left(t\right)$ the star-planet separation, $\alpha_\star=\alpha_\star\!\left( \phi, \theta, t \right)$ the local stellar altitude, and $T_0$ the minimum temperature parameter. The stellar altitude $\alpha_\star$ is, relative to the sub-stellar point pointed to by $\hat{r}_\star = \phi_\star\hat{\phi} + \theta_\star\hat{\theta}$ and the unit normal $\hat{n}$ at the position, 
\begin{equation}\label{eq:stellar_altitude}
\cos\alpha_\star = 
\begin{cases}
\hat{n}\cdot\hat{r}_\star, &\hat{n}\cdot\hat{r}_\star \geq 0 \\
0, &\hat{n}\cdot\hat{r}_\star < 0
\end{cases}
\end{equation}
where $\hat{n}\cdot\hat{r}_\star \rightarrow \hat{r}\cdot\hat{r}_\star = \cos\theta\cos\theta_\star \left[\cos\left(\phi-\phi_\star\right)-1\right] + \cos\left(\theta-\theta_\star\right)$ for spherical planets, and the sub-stellar point is given by\footnote{For the equivalent derivation in inertial Cartesian coordinates, see \S 3 in \citet{Dobrovolskis2013}.} 
\begin{align}\label{eq:substellar_point}
\begin{split}
\phi_\star =\ &\phi_\star\!\left(t_0\right) - \Big\{\omega_\mathrm{rot}\left(t-t_0\right) + \\ &\arctan\left\{\cos\obliq \tan\left[\left(\nu-\axaz\right)-\nu\!\left(t_0\right)\right]\right\} \Big\} \\
\theta_\star =\ &\sin^{-1}\left[\sin\obliq \cos\left(\nu-\axaz\right)\right].
\end{split}
\end{align}
For tidally distorted planets, the unit normal $\hat{n}$ no longer matches the position unit vector $\hat{r}$; \S \ref{sec:model:tidal} covers this case.

Finally, the change in temperature of each cell in time is calculated as
\begin{equation}\label{eq:T}
    \dot{T}\!\left( \phi, \theta, t  \right) = \frac{T_{\textrm{eq}}}{4 \tau_{\textrm{rad}}} \left\{ 1 - \left[\frac{T\!\left( \phi, \theta, t  \right)}{T_{\textrm{eq}}}\right]^4 \right\}\,,
\end{equation}
which is a differential equation we can evaluate numerically for sufficiently small timesteps. We choose to divide each orbit into 200 timesteps.

\subsection{Generating Observables}\label{sec:model:observables}
Given a temperature map $T\!\left(\phi,\theta,t\right)$, we solve for the corresponding planet-star flux contrast via
\begin{equation}\label{eq:contrast}
\bar{F}\!\left(t\right) = \frac{1}{\pi}\left(\frac{R_\mathrm{p}}{R_\star}\right)^2 \frac{\iiint w\,B_\lambda\!\left(T\right) V \,d\lambda\,d\theta\,d\phi}{\int w\,B_\lambda\!\left(T_\star\right) \,d\lambda}
\end{equation}
where $B_\lambda\!\left(T\right)$ is the specific blackbody intensity at a wavelength $\lambda$ and temperature $T$, $w=w\!\left(
\lambda\right)$ the weighted response of the instrumental bandpass at $\lambda$, and $V=V\!\left(\phi,\theta,t\right)$ is the component of the normal vectors of the cells along the line of sight, given by
\begin{equation}\label{eq:visibility}
V = 
\begin{cases}
\hat{n}\cdot\hat{r}_\mathrm{obs}, &\hat{n}\cdot\hat{r}_\mathrm{obs} \geq 0 \\
0, &\hat{n}\cdot\hat{r}_\mathrm{obs} < 0
\end{cases}
\end{equation}
where the sub-observer point is given by
\begin{align}\label{eq:subobserver_point}
\begin{split}
\phi_\mathrm{obs} &= \phi_\star\!\left(t_\mathrm{ecl}\right) - \omega_\mathrm{rot}\left(t-t_\mathrm{ecl}\right) \\
\theta_\mathrm{obs} &= \sin^{-1}\left[\sin\obliq \cos\!\left(\nu_\mathrm{ecl}-\axaz\right)\right]
\end{split}
\end{align}
and $\nu_\mathrm{ecl}=3\pi/2-\varpi$ is the true anomaly during secondary eclipse (i.e.\ at time $t_\mathrm{ecl}$), and for spherical planets, $\hat{n}\cdot\hat{r}_\mathrm{obs} = \cos\theta\cos\theta_\mathrm{obs} \left[\cos\left(\phi-\phi_\mathrm{obs}\right)-1\right] + \cos\left(\theta-\theta_\mathrm{obs}\right)$. However, as with the stellar altitude, for non-spherical planets the calculation is more involved, as we discuss in the following section.

\subsection{A Simple Model of Tidal Distortion}\label{sec:model:tidal}
A tidally distorted planet can have a significantly aspherical shape. We first outlined a model of tidal asphericity in Appendix A of \citet{ada18b}. Here we adopt that work's primary assumption to model the distorted shape as a prolate spheroid, with the long axis displaced clockwise in the orbital plane from the star-planet line by some lag angle $\lambda \equiv \cos^{-1}\!\left(\hat{r}_\star \cdot \hat{r}_\ell\right)$. We will assume this lag angle is zero for our analysis, but include it for completeness. The lengths of the long and short axes are dictated by the gravitational potential of the star-planet system. Consider a coordinate system where $\hat{z}$ points from the planet center along the long axis, $\hat{y}$ points along the orbit normal, and $\hat{x}$ points along the short axis according to a right-handed coordinate system. Then the potential becomes
\begin{align}\label{eq:potential}
\begin{split}
\Phi \!\left(\vec{r}\right) = -\frac{GM_\star}{a} \Bigg\{&\left[\left(\frac{z}{a}+\frac{\xi}{1+\xi}\right)^2 + \left(\frac{x^2+y^2}{a^2}\right)\right]^{-1/2} \\ + &\left[\left(\frac{z}{a}+\frac{1}{1+\xi}\right)^2 + \left(\frac{x^2+y^2}{a^2}\right)\right]^{-1/2} \\ + &\frac{1+\xi}{2} \left(\frac{x^2+z^2}{a^2}\right)\Bigg\}
\end{split}
\end{align}
where $a$ is the orbital semi-major axis and $\xi$ is the planet-star mass ratio $M_\mathrm{p} / M_\star$. We then fit the cross-sectional area along $\hat{x}$ and $\hat{y}$ to the observed transit depth to get the long and short axis lengths\footnote{In general the solution yields unequal extents along $\hat{x}$ and $\hat{y}$, while the prolate spheroid assumption implies they are equal. For \WASP{}, the differences are minor enough that the corresponding differences in calculated ellipticity are $\lesssim$ 0.01.}.

For an ellipsoid the area of each cell is position-dependent, and will therefore affect the surface area over which it radiates. Once we have the planetary semi-major and semi-minor extents (defined as one-half of the long and short axes, respectively), denoted $A_\mathrm{p}$ and $B_\mathrm{p}$, we can calculate the areas of individual cells. We adapt the result from Equation A7 of \citet{ada18b} for the area of a cell spanning longitudes $\phi_i$--$\phi_j$ and latitudes $\theta_i$--$\theta_j$, now with a more complicated relationship between the Cartesian coordinates for the planet:
\begin{align}\label{eq:cell_area}
\begin{split}
S_{ij} &= \, B^2_\mathrm{p} \int_{\phi_i}^{\phi_j} \!\int_{\theta_i}^{\theta_j}\! \cos\theta \, \Big\{ 1 \\ &+ \left(\chi^2-1\right) \left[\frac{1}{4}f\!\left(\phi,\theta\right)+2\left(\hat{r}\cdot\hat{r}_\ell\right)\right] \\ &+ \left(\chi^2-1\right)^2 \left(\hat{r}\cdot\hat{r}_\ell\right)^2 \Big\}^{1/2} \,d\theta\,d\phi,
\end{split}
\end{align}
where $\chi \equiv A_\mathrm{p}/B_\mathrm{p}$ is the axis ratio and
\begin{align}\label{eq:area_term}
\begin{split}
f\!\left(\phi,\theta\right) &= \left\{\sin\theta\cos\theta_\ell\left[\cos\left(\phi-\phi_\ell\right)-1\right] - \sin\left(\theta-\theta_\ell\right)\right\}^2 \\ &+ \cos^2\theta_\ell \left[\sin\left(\phi-\phi_\ell\right)+1\right].
\end{split}
\end{align}

\begin{figure*}[htb!]
\begin{center}
\begin{tabular}{ccc}
\textbf{\HD{}} & \textbf{\WASP{}} & \textbf{\COROT{}} \\
\includegraphics[height=5.5cm]{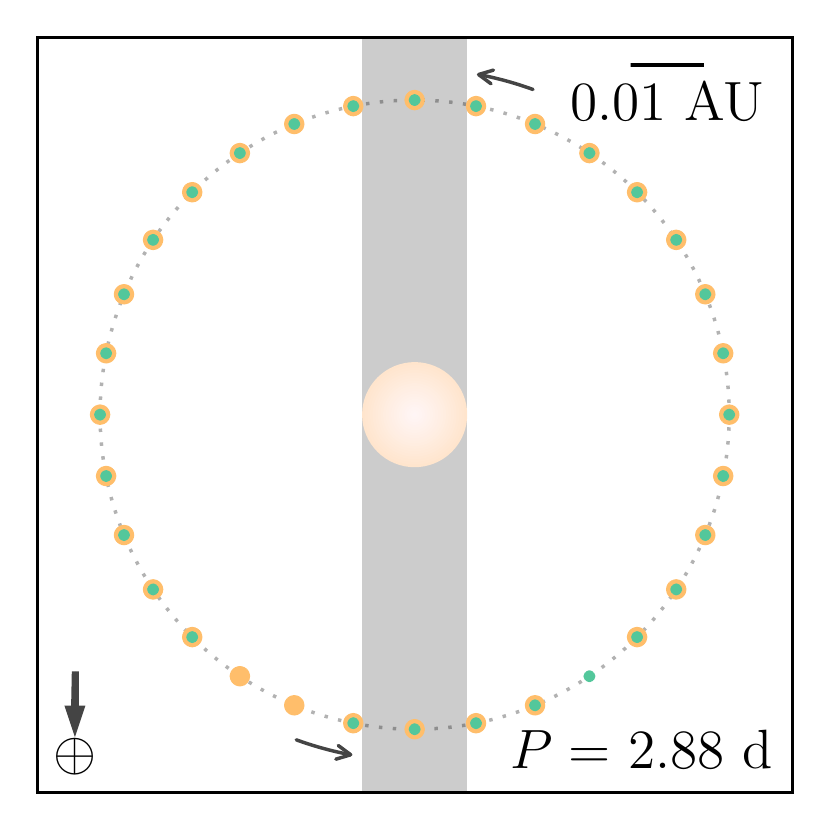} &
\includegraphics[height=5.5cm]{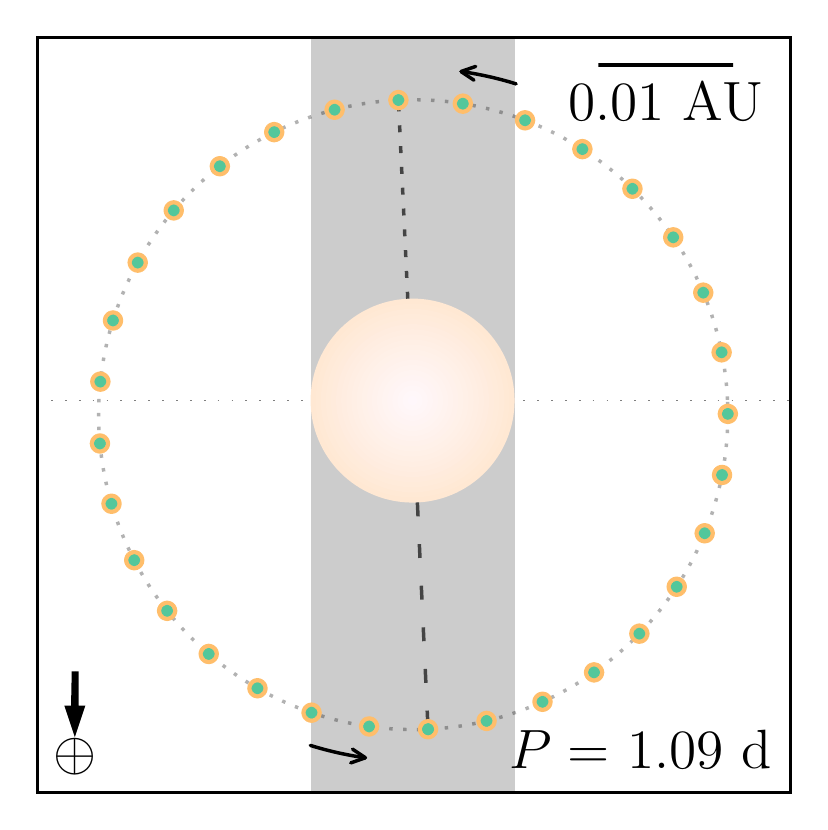} &
\includegraphics[height=5.5cm]{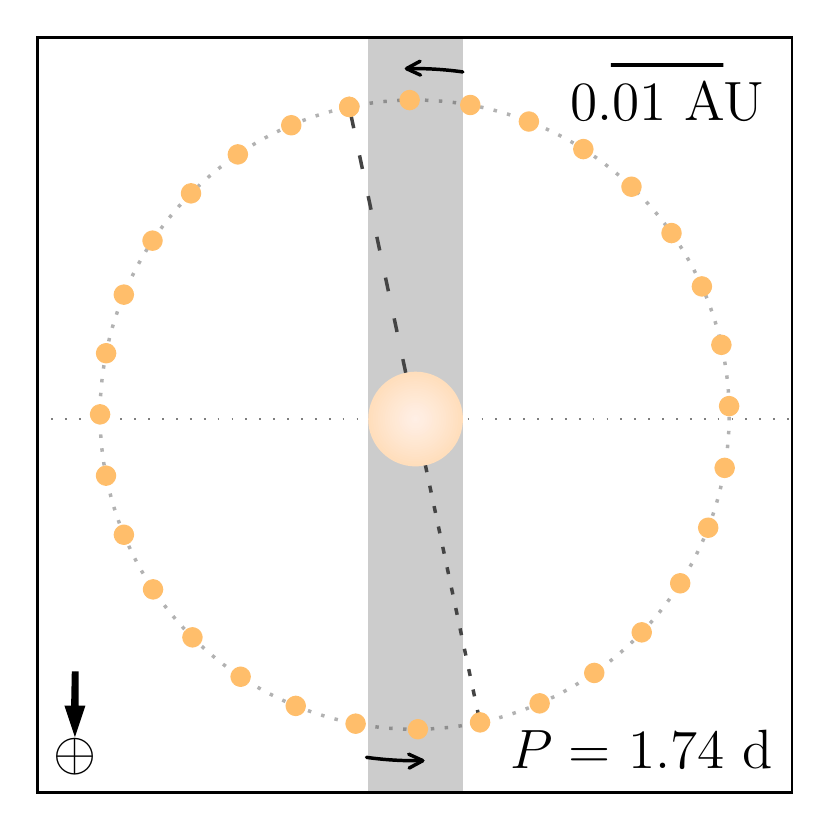}
\end{tabular}
\caption{Top-down orbital diagrams of \HD{} (left), \WASP{} (center), and \COROT{} (right). The stars are drawn to scale with respect to the orbits, and the concentric colored rings show a binned coverage of each phase curve, with 3.6 $\mu$m in green and 4.5 $\mu$m in yellow. The orbits of \WASP{} and \COROT{} are slightly eccentric; the positions of periastron and apastron are shown by the dashed lines with small and large lengths/spacings, respectively.}
\label{fig:planet_orbits}
\end{center}
\end{figure*}

A second effect of the non-spherical shape is a change in the stellar altitude as a function longitude and latitude. To quantify the change we first write the function representing the shape of the prolate ellipsoid:
\begin{equation}\label{eq:ellipsoid_function}
f\!\left(x,y,z\right) = \frac{x^2+y^2}{B^2_\mathrm{p}} + \frac{z^2}{A^2_\mathrm{p}}.
\end{equation}
We need a way of expressing these ellipsoidal coordinates in the oblique coordinates (i.e. with respect to the latitude/longitude coordinates defined by the rotation). To do this we note that we can express the oblique positions of both the sub-stellar point, given by equation \ref{eq:substellar_point}, and the extreme point of the planet along the long axis, given by
\begin{align}\label{eq:ellipsoidal_point}
\begin{split}
\phi_\ell &= \phi_\star\!\left(\nu\rightarrow\nu-\lambda\right) \\
\theta_\ell &= \theta_\star\!\left(\nu\rightarrow\nu-\lambda\right). \\
\end{split}
\end{align}
If the lag angle $\lambda>0$, then we can construct our ellipsoidal Cartesian unit vectors entirely with respect to the unit position vectors for these two points:
\begin{align}\label{eq:ellipsoidal_units}
\begin{split}
\hat{x} &= \frac{\hat{r}_\star - \cos\lambda\hat{r}_\ell}{\sin\lambda} \\
\hat{y} &= \frac{\hat{r}_\ell\times\hat{r}_\star}{\sin\lambda} \\
\hat{z} &= \hat{r}_\ell.
\end{split}
\end{align}
From this the normal unit vector at a given point on the surface is given by
\begin{align}\label{eq:ellipsoid_normal}
\begin{split}
\hat{n} &= \vec{\nabla}\!f / \left\lVert \vec{\nabla}\!f \right\rVert \\
&= \left\{1 + g^{-1}\!\left[\chi,\left(\hat{r}\cdot\hat{r}_\ell\right)\right]\right\}^{-1/2} \left[\left(\hat{r}\cdot\hat{x}\right)\hat{x} + \left(\hat{r}\cdot\hat{y}\right)\hat{y}\right] \\
&+ \left\{1 + g\!\left[\chi,\left(\hat{r}\cdot\hat{r}_\ell\right)\right]\right\}^{-1/2} \hat{z}
\end{split}
\end{align}
where
\begin{equation}
g\!\left[\chi,\left(\hat{r}\cdot\hat{r}_\ell\right)\right] \equiv  \chi^{2} \left[\frac{1-\left(\hat{r}\cdot\hat{r}_\ell\right)^2}{\left(\hat{r}\cdot\hat{r}_\ell\right)^2}\right]
\end{equation}
Then the cosines of the stellar altitude and the visibility are given by the component of $\hat{n}$ along the instellation and observer lines, as in Equations \ref{eq:stellar_altitude} and \ref{eq:visibility}.
\begin{align}\label{eq:ellipsoid_cosalpha}
\begin{split}
\cos\alpha_\ell &= \hat{n} \cdot \hat{r}_\star \\
&= \sin\lambda \left(1 + g^{-1}\right)^{-1/2} \left[\hat{r}\cdot\hat{r}_\star-\cos\lambda\left(\hat{r}\cdot\hat{r}_\ell\right)\right] \\
&+ \cos\lambda \left(1 + g\right)^{-1/2} \left(\hat{r}\cdot\hat{r}_\ell\right)
\end{split}
\end{align}
and the component along the observer line is
\begin{align}\label{eq:ellipsoid_cosobs}
\begin{split}
V &= \hat{n}\cdot\hat{r}_\mathrm{obs} \\
&= \frac{\cos\left(\nu-\nu_\mathrm{ecl}\right) - \cos\left[\left(\nu-\lambda\right)-\nu_\mathrm{ecl}\right]}{\sin\lambda} \left(1 + g^{-1}\right)^{-1/2} \\
&\times \left[\hat{r}\cdot\hat{r}_\star-\cos\lambda\left(\hat{r}\cdot\hat{r}_\ell\right)\right] \\
&+ \cos\left[\left(\nu-\lambda\right)-\nu_\mathrm{ecl}\right] \left(1 + g\right)^{-1/2} \left(\hat{r}\cdot\hat{r}_\ell\right).
\end{split}
\end{align}

\section{Fit Methods and Results}\label{sec:results}

\begin{deluxetable*}{cccccc}
\tabletypesize{\footnotesize}
\tablewidth{0pt}
\tablecaption{Best-Fit Parameters from Radiative Model}
\tablehead{\colhead{} & \multicolumn{2}{c}{\HD{}} & \multicolumn{2}{c}{\WASP{}} & \colhead{\COROT{}} \\ \colhead{Parameter} & \colhead{3.6 $\mu$m} & \colhead{4.5 $\mu$m} & \colhead{3.6 $\mu$m} & \colhead{4.5 $\mu$m} & \colhead{4.5 $\mu$m}}
\startdata
\sidehead{Sub-synchronous Rotation}
$\tau_{\mathrm{rad}}$ (hr) & $7.7^{+10.8}_{-4.6}$ & $2.4^{+24.0}_{-1.3}$ & $0.01^{+0.01}_{-0.01}$ & $65.5^{+24.3}_{-0.3}$ & $23.9^{+2.5}_{-3.1}$ \\
$T_0$ (K) & $877^{+388}_{-631}$ & $1388^{+155}_{-158}$ & $427^{+166}_{-N/A}$ & $2458^{+84}_{-141}$ & $92^{+209}_{-52}$ \\
$A$ & $<0.12$ & $0.64^{+0.09}_{-0.05}$ & $<0.06$ & $0.36^{+0.09}_{-0.35}$ & $<0.03$ \\ $\obliq$ ($^\circ$) & $93.9^{+45.1}_{-32.5}$ & $4.2^{+47.7}_{-1.9}$ & $87.9^{+0.6}_{-8.5}$ & $91.2^{+70.1}_{-3.1}$ & $45.8\pm1.4$ \\
$\axaz$ ($^\circ$) & $10.6^{+143.7}_{-170.42}$ & $-29.0^{+113.0}_{-149.2}$ & $-39.4^{+4.4}_{-0.9}$ & $-73.0^{+21.2}_{-58.6}$ & $-82.5^{+10.8}_{-7.1}$ \\
\sidehead{Free Rotation}
$P_{\mathrm{rot}}/P_{\mathrm{orb}}$ & $1.09\pm0.06$ & $0.39^{+0.08}_{-0.21}$ & $0.77^{+0.14}_{-0.04}$ & $0.94\pm0.01$ & $1.13^{+0.04}_{-0.02}$ \\
$\tau_{\mathrm{rad}}$ (hr) & $97.2^{+5.1}_{-0.2}$ & $3.7^{+0.4}_{-2.2}$ & $12.7^{+32.2}_{-2.7}$ & $40.0^{+8.1}_{-3.4}$ & $16.9^{+4.7}_{-4.4}$ \\
$T_0$ (K) & $1172^{+334}_{-111}$ & $1223^{+165}_{-894}$ & $1567^{+134}_{-63}$ & $2204^{+40}_{-14}$ & $102^{+219}_{-75}$ \\
$A$ & $0.01^{+0.39}_{-0.01}$ & $0.44^{+0.05}_{-0.34}$ & $0.09^{+0.02}_{-0.08}$ & $<0.03$ & $<0.03$ \\
$\obliq$ ($^\circ$) & $4.2^{+57.2}_{-2.7}$ & $36.4^{+28.3}_{-3.4}$ & $0.8^{+6.0}_{-0.4}$ & $21.1^{+8.5}_{-10.1}$ & $2.0^{+8.1}_{-1.2}$\\
$\axaz$ ($^\circ$) & $-4.7^{+7.3}_{-30.2}$ & $69.8^{+3.0}_{-29.5}$ & $-67.0^{+123.4}_{-N/A}$ & $46.6^{+58.3}_{-12.1}$ & $-32.3^{+47.1}_{-20.8}$ \\
\enddata
\tablecomments{The parameter values from our blackbody model returning the most favorable likelihood from MCMC algorithms. Uncertainties listed are 1-$\sigma$ ranges of a Metropolis-Hastings algorithm walk around the region of most favorable likelihood in parameter space. Upper limits imply the best-fit values are zero, with a 1-$\sigma$ uncertainty given by the upper limit.}
\label{table:fits}
\end{deluxetable*}

\begin{figure*}[htb!]
\begin{center}
\textbf{\HD{}}\par\medskip
\begin{tabular}{rcc}
\includegraphics[width=7.5cm]{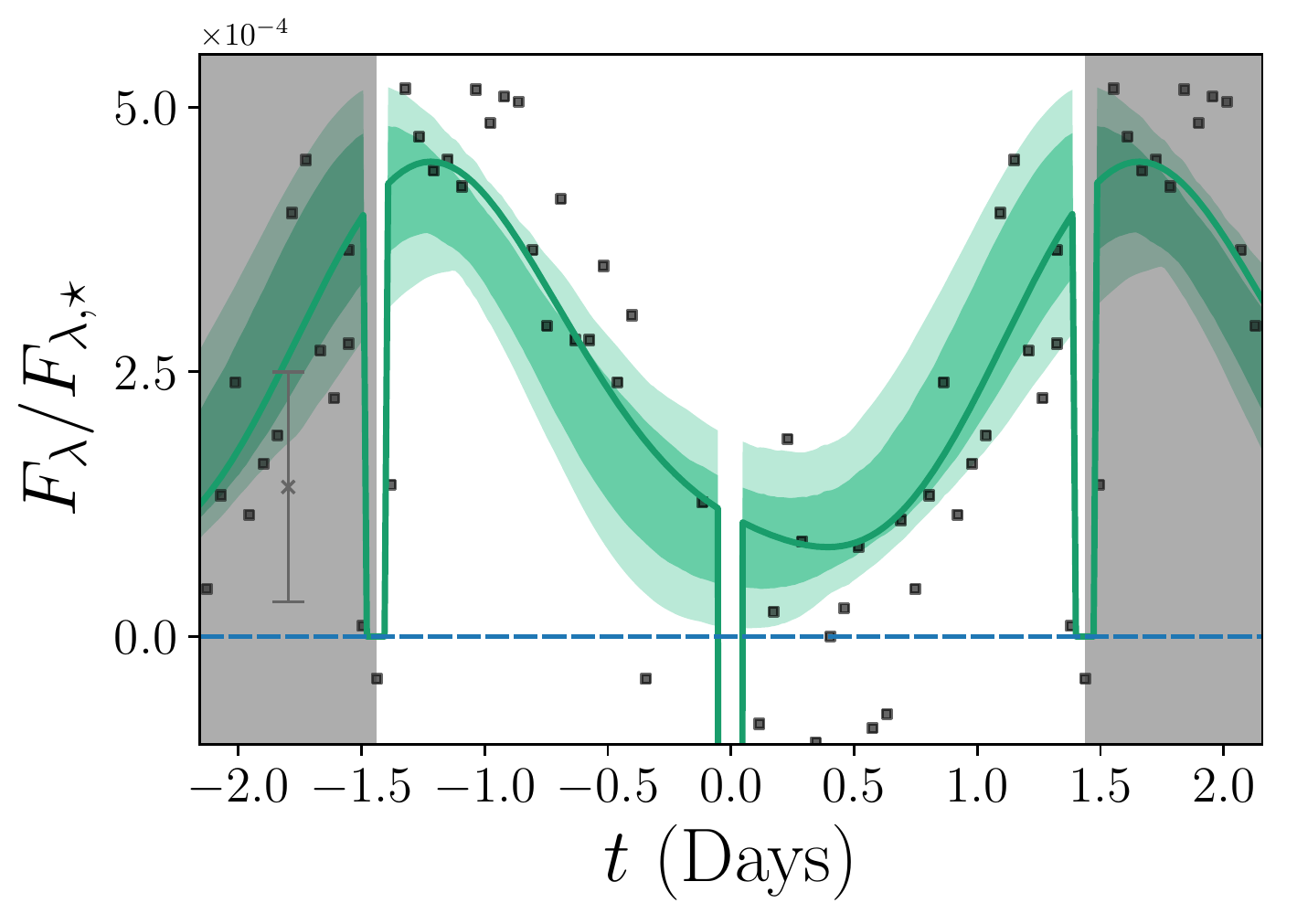} &
\raisebox{0.25\height}{\includegraphics[width=5.5cm]{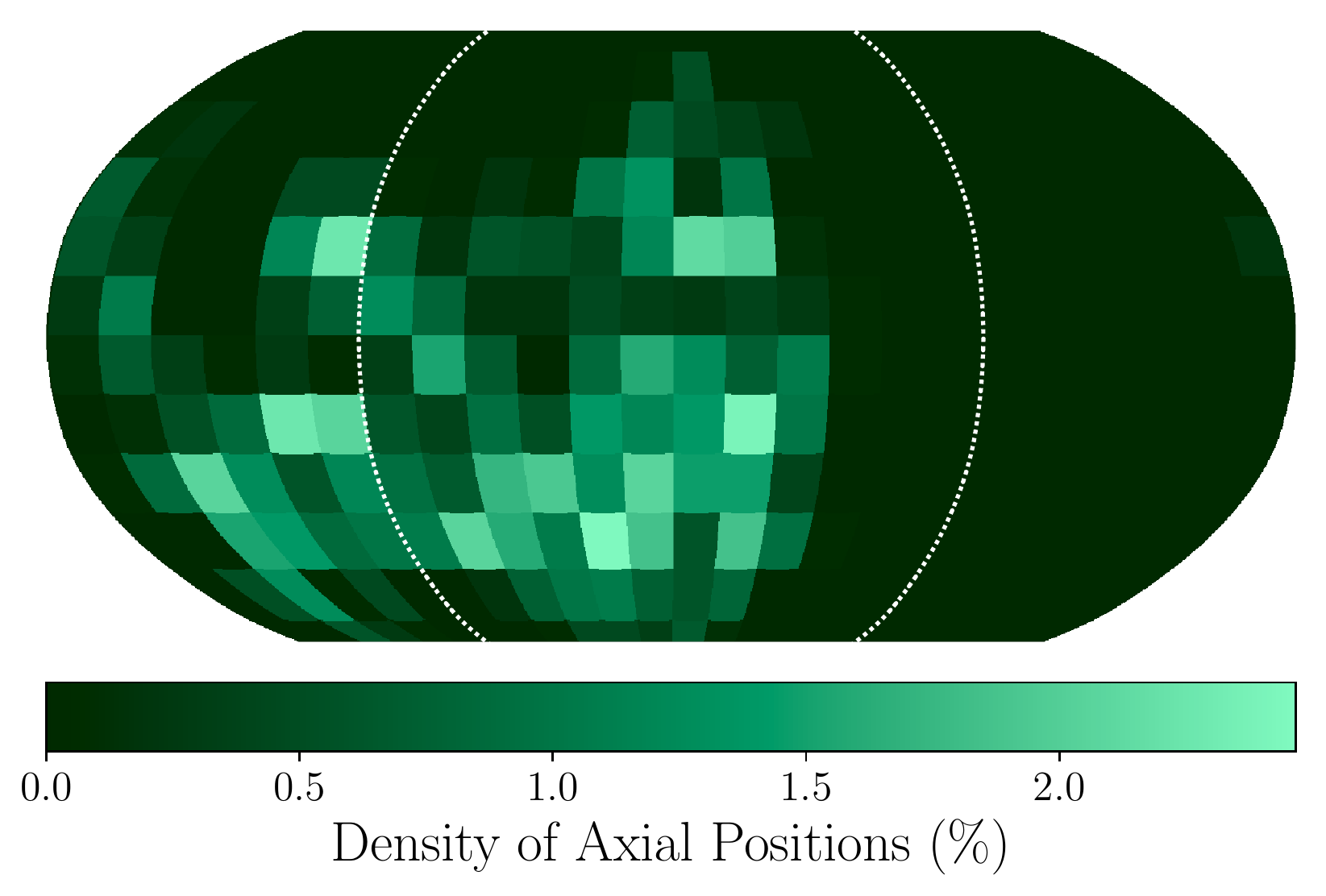}} &
\raisebox{0.45\height}{\includegraphics[width=3.5cm]{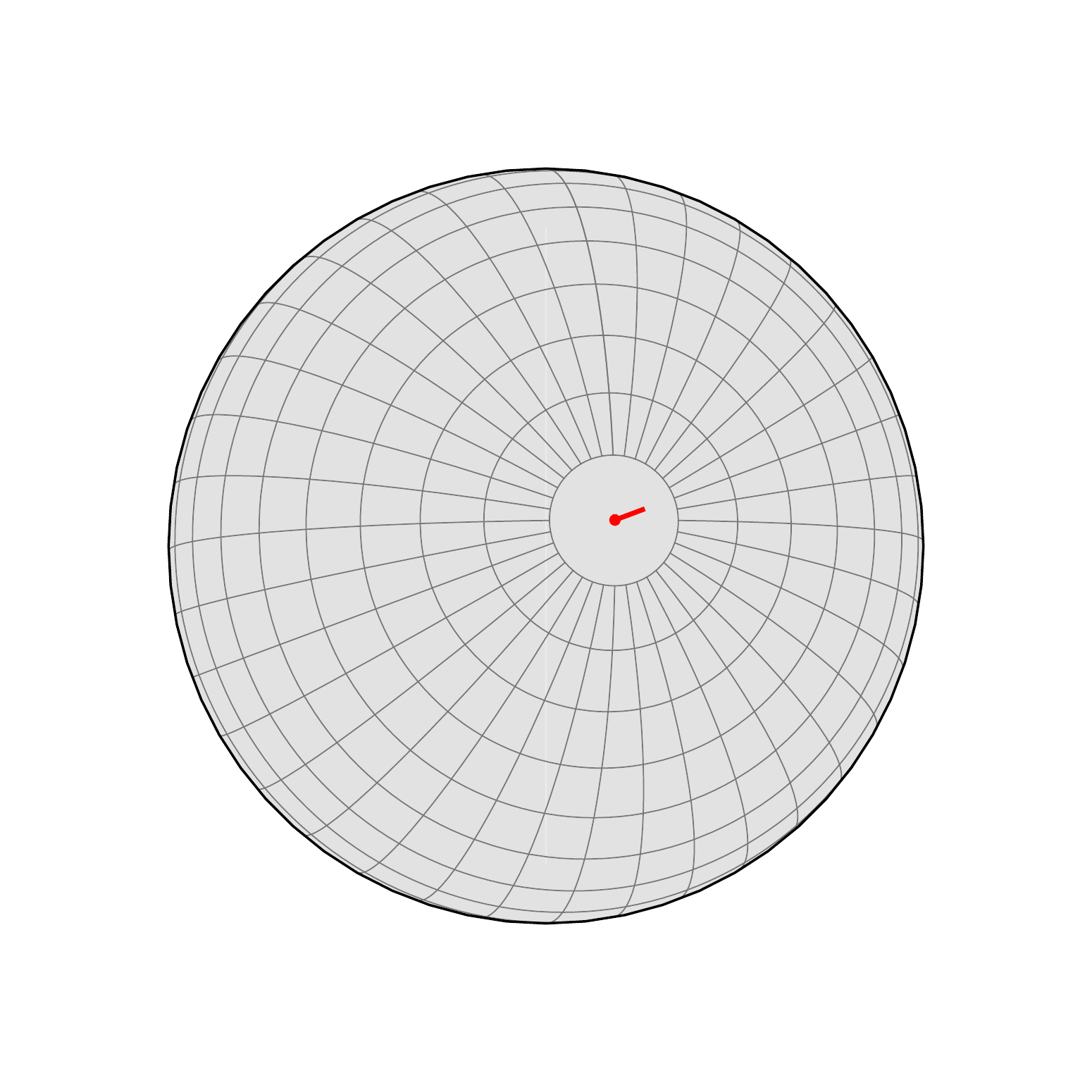}} \\
\includegraphics[width=7.5cm]{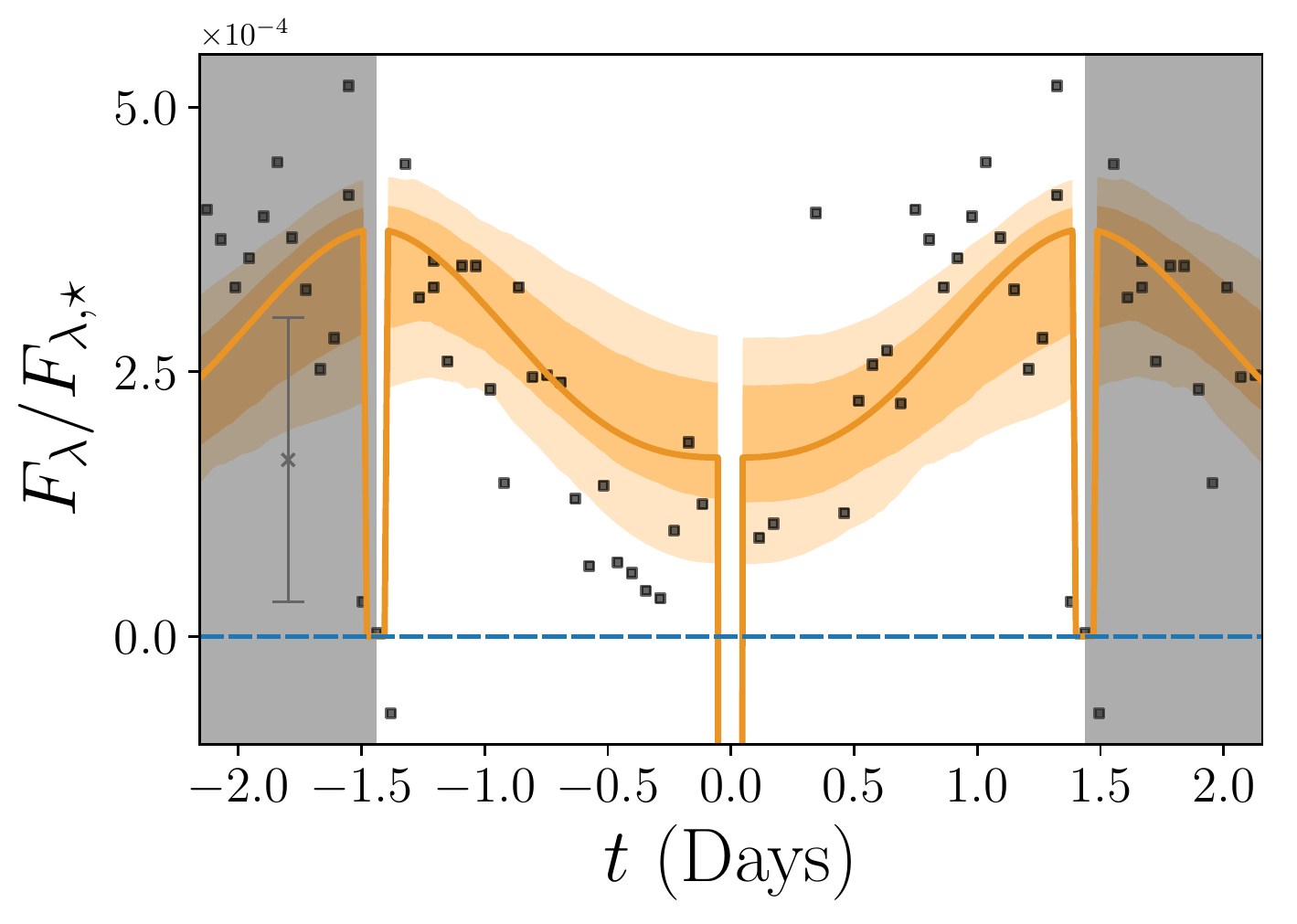} &
\raisebox{0.25\height}{\includegraphics[width=5.5cm]{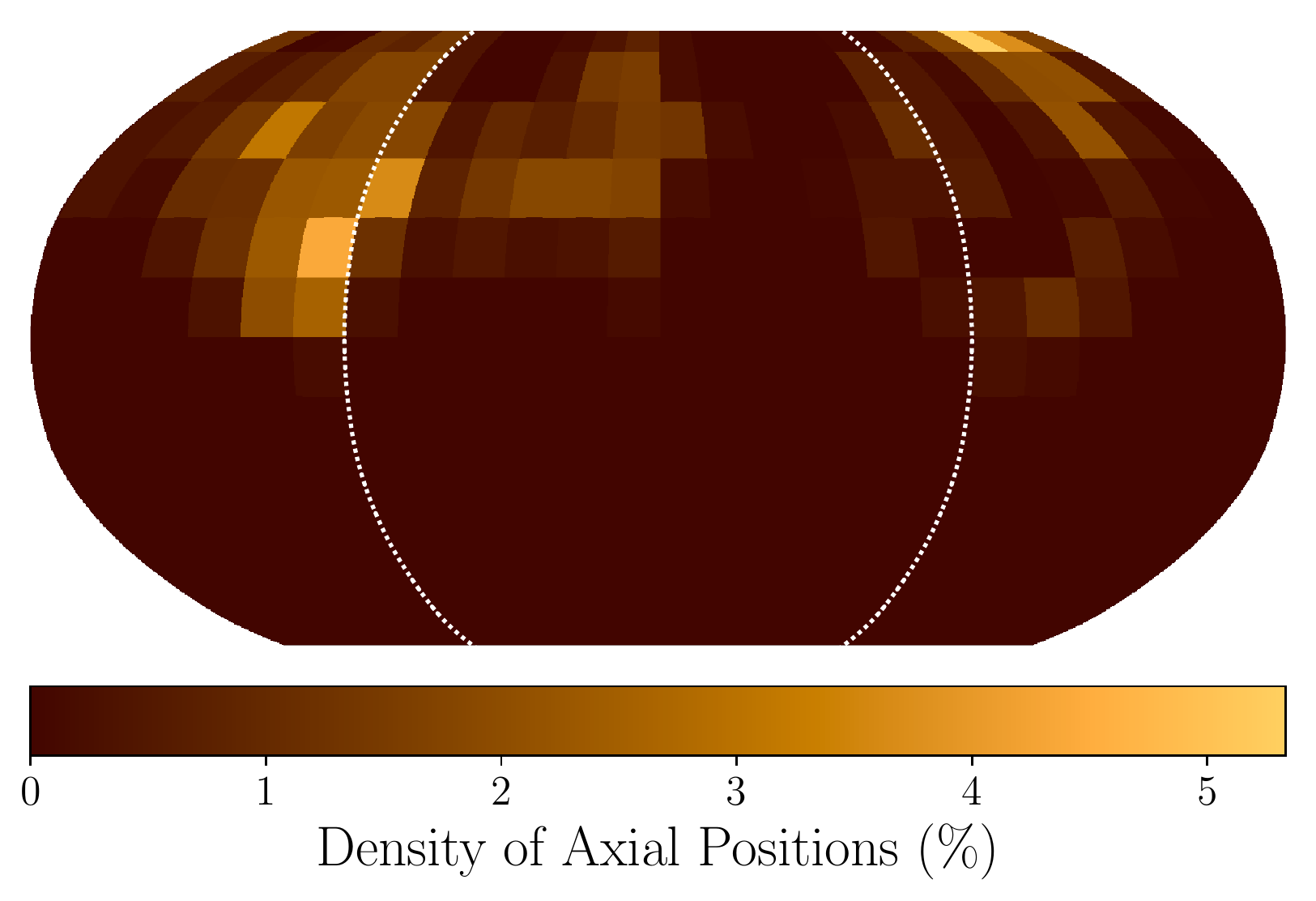}} &
\raisebox{0.45\height}{\includegraphics[width=3.5cm]{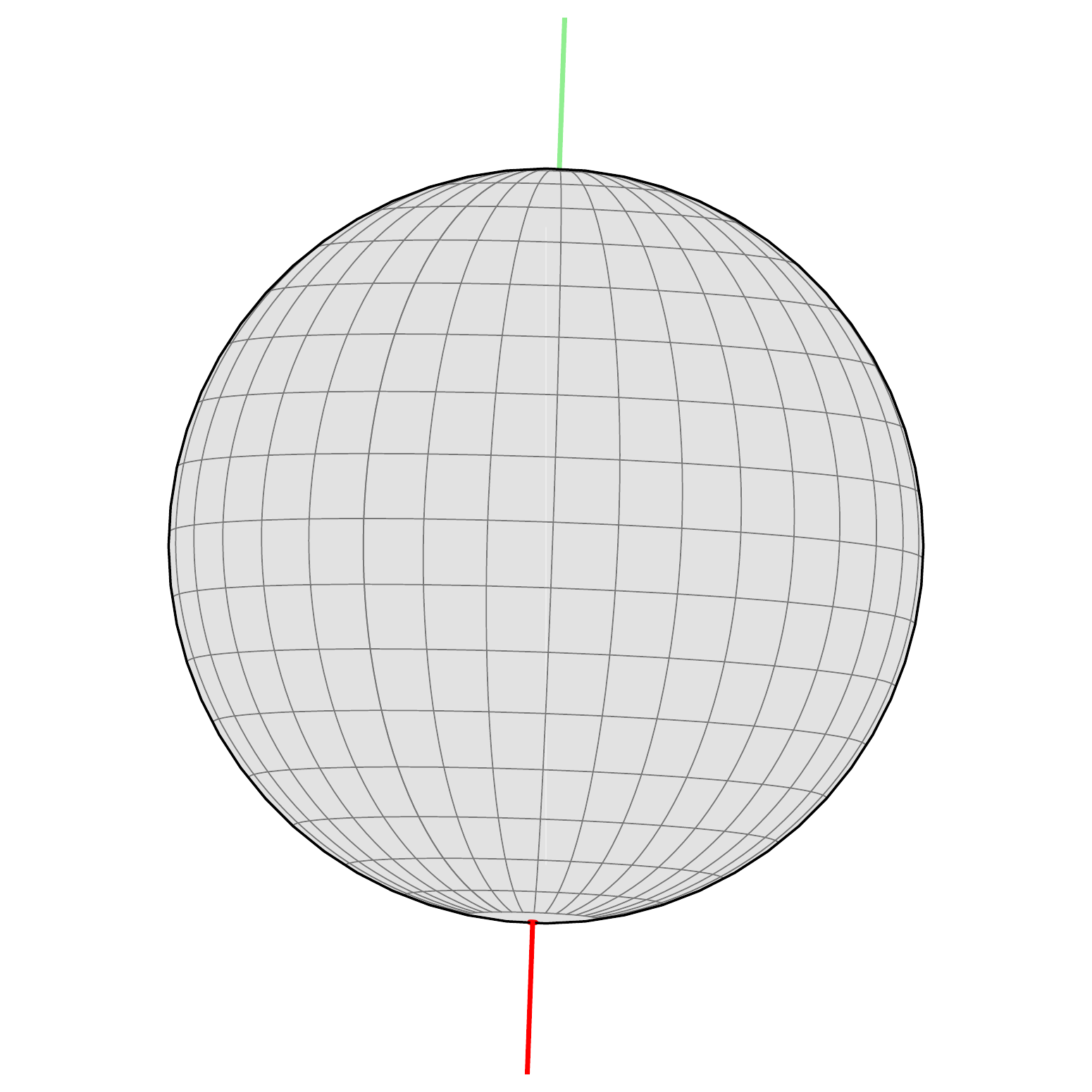}} \\
\end{tabular}
\caption{\emph{Left}: the best-fit oblique thermal models for \HD{} (the sub-synchronous case in Table \ref{table:fits}), in the 3.6 (top) and 4.5 (bottom) $\mu$m channels. The colored lines represent the best-fit light curve, with the shaded regions representing the 1- and 2-$\sigma$ uncertainty bounds (darker and lighter, respectively). The black points represent the binned data from \citet{zha18}. \emph{Center}: the density of explored axial orientations from the uncertainty calculations in the MCMC routine. The observer-facing hemisphere is framed by the dotted white lines. \emph{Right}: globes showing the best-fit orientation of the spin axes for each band, as viewed along our line of sight.}
\label{fig:HD149026b_curves}
\end{center}
\end{figure*}

\begin{figure*}[htb!]
\begin{center}
\textbf{\WASP{}}\par\medskip
\begin{tabular}{rcc}
\includegraphics[width=7.5cm]{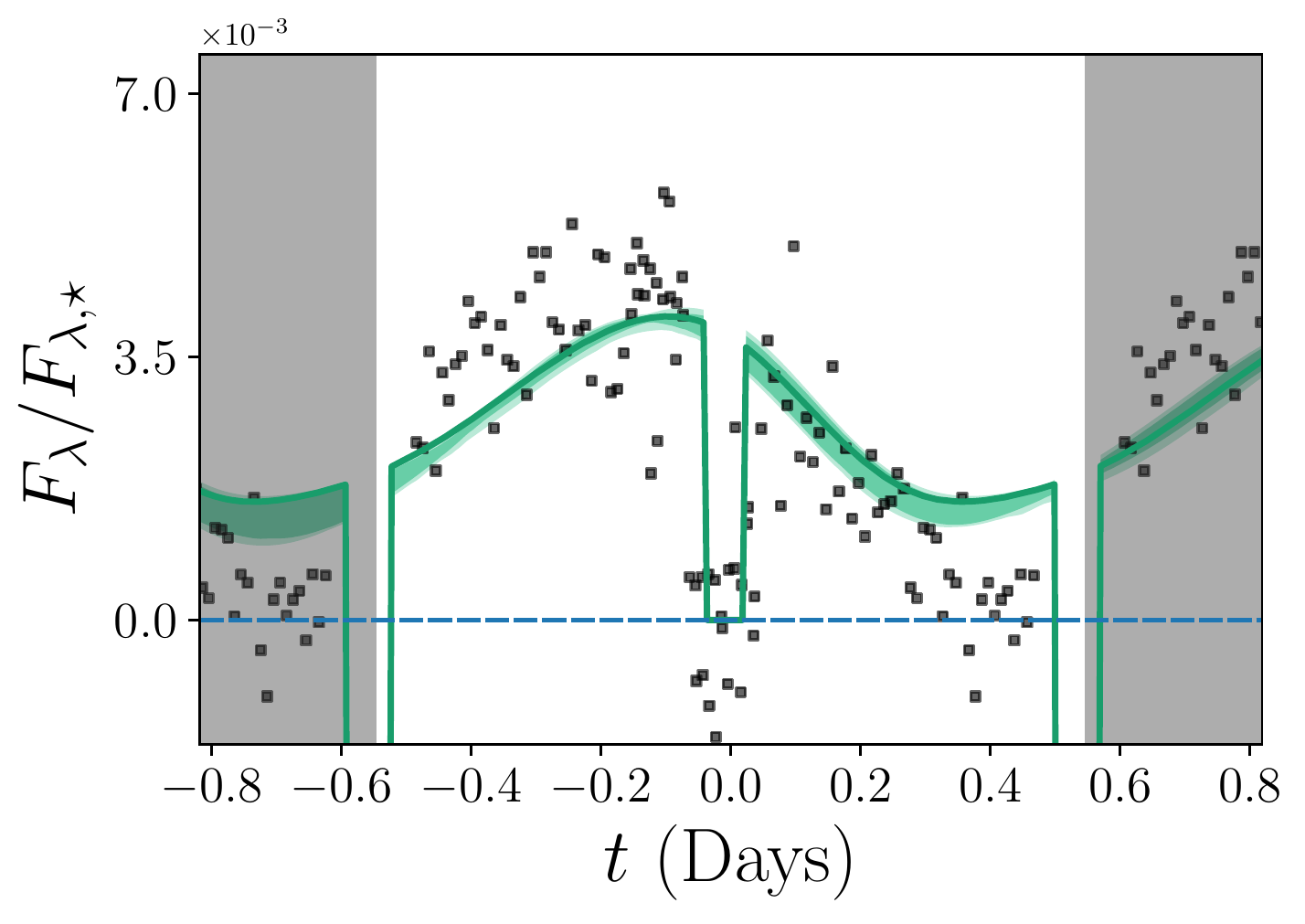} &
\raisebox{0.25\height}{\includegraphics[width=5.5cm]{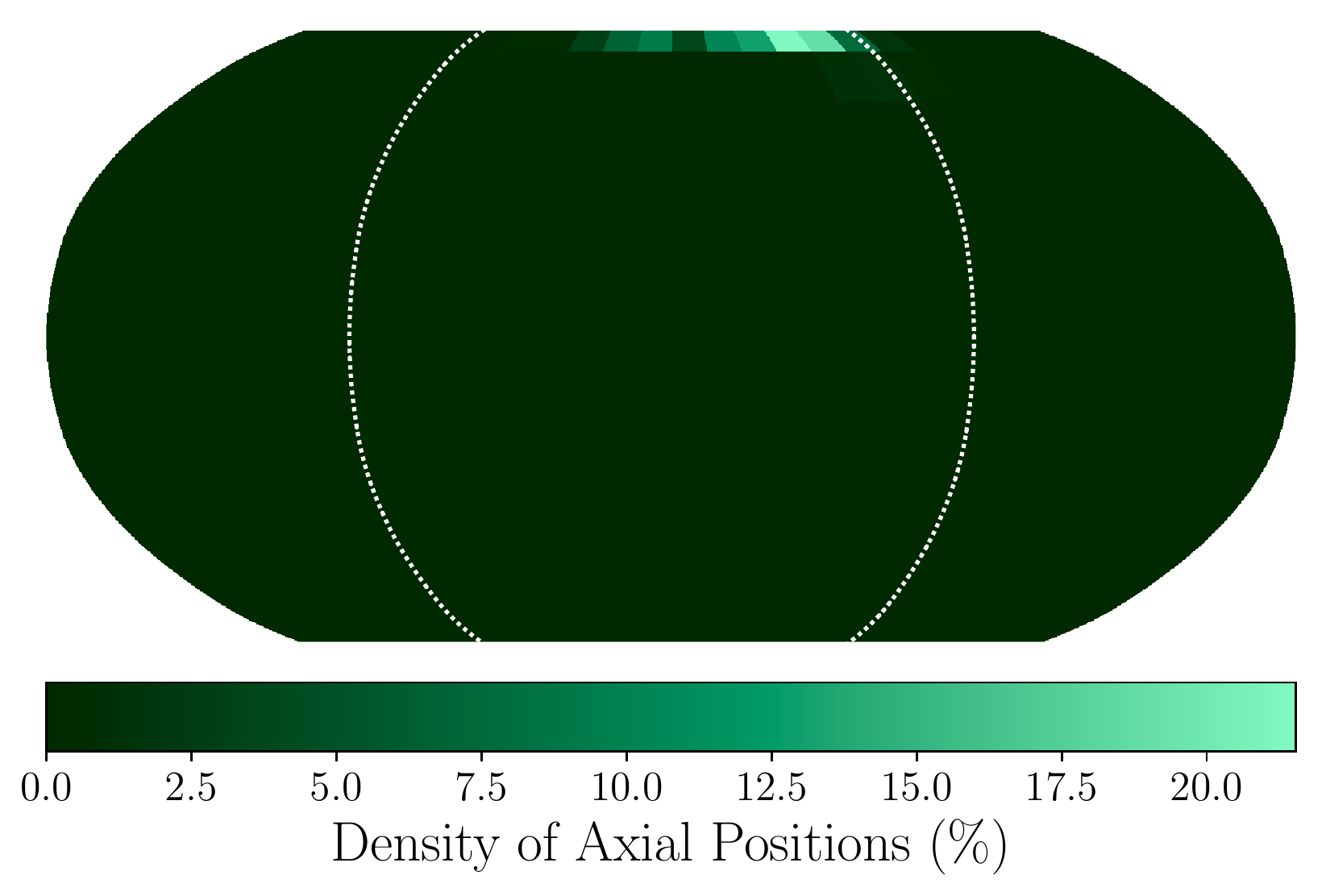}} &
\raisebox{0.45\height}{\includegraphics[width=3.5cm]{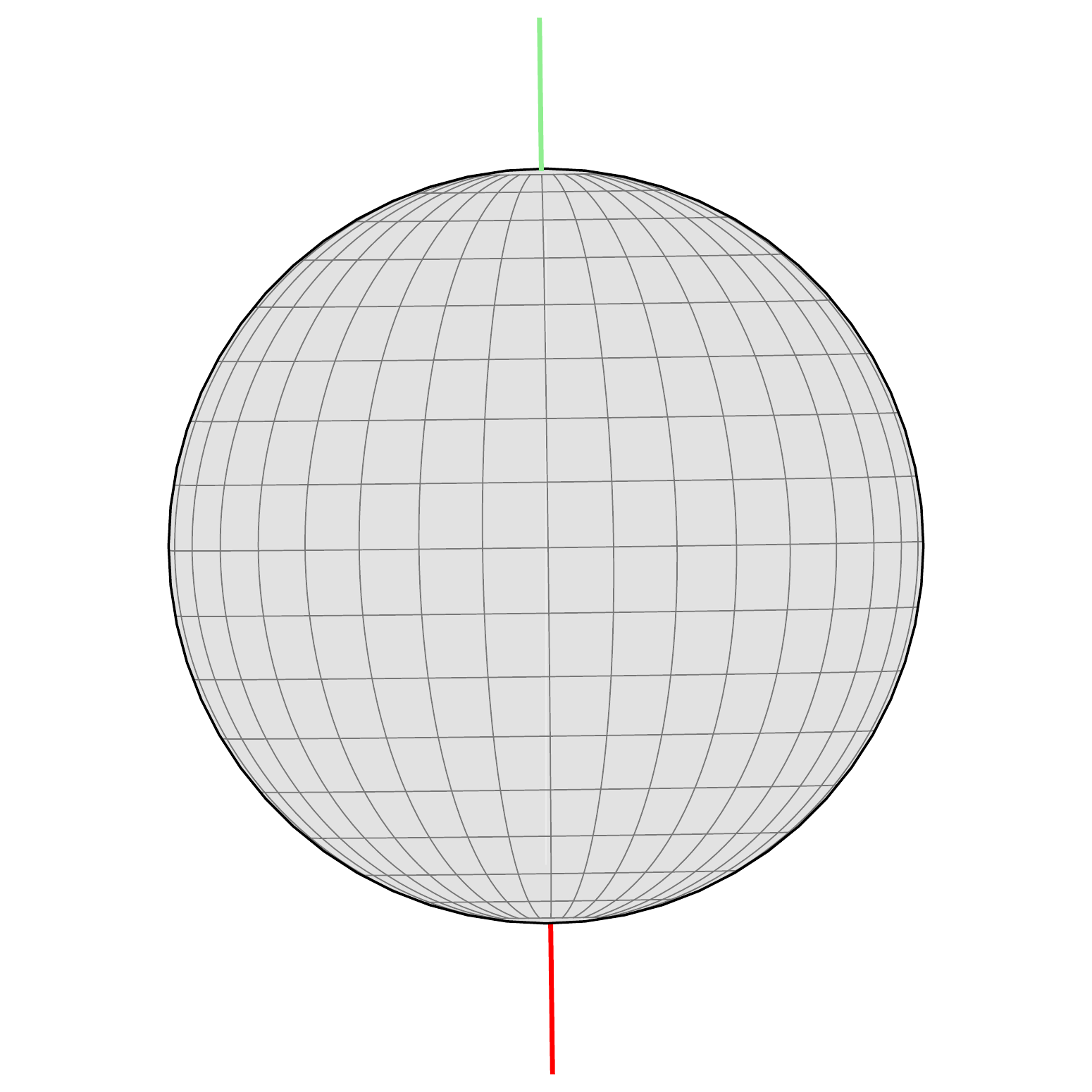}} \\
\includegraphics[width=7.5cm]{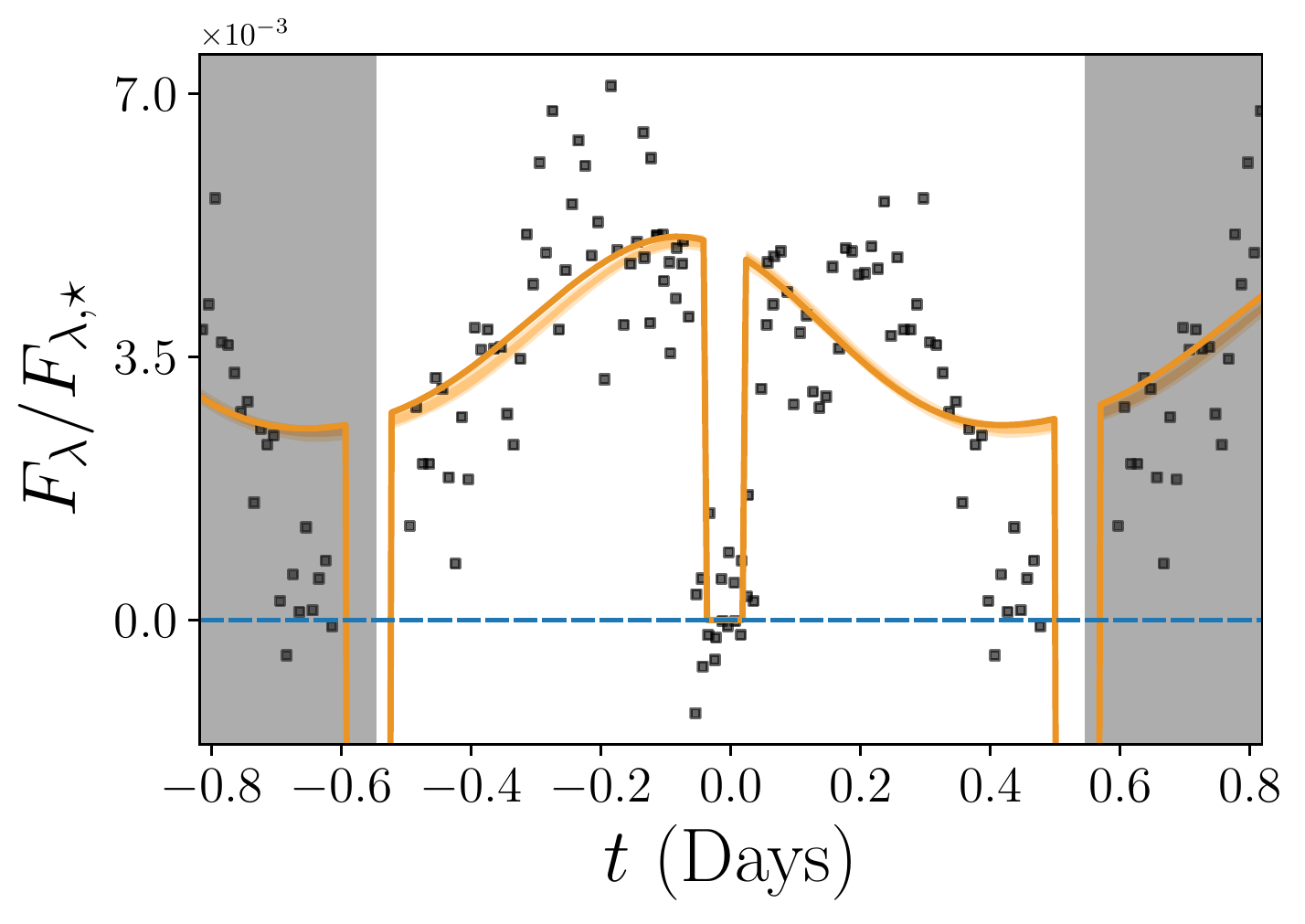} &
\raisebox{0.25\height}{\includegraphics[width=5.5cm]{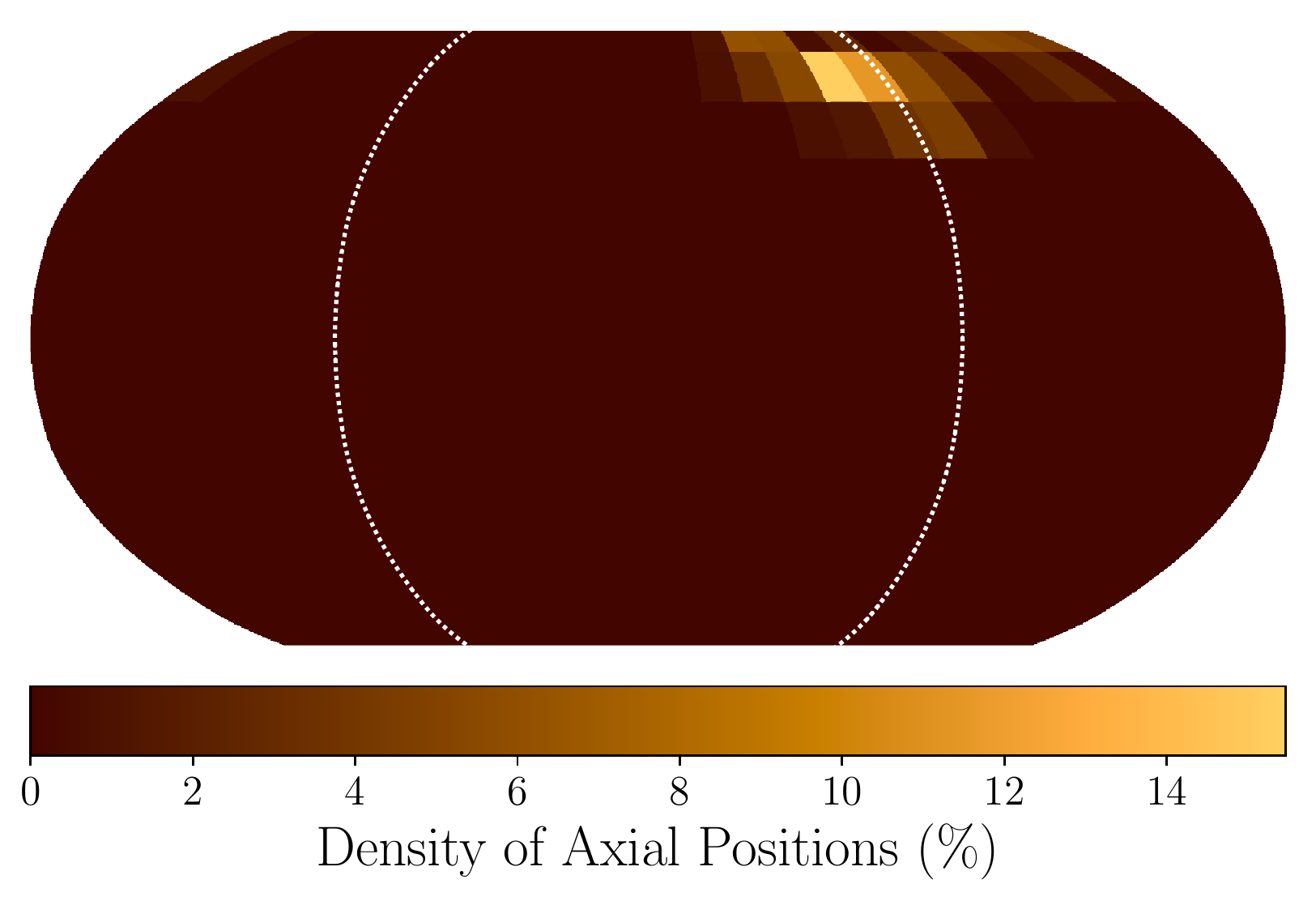}} &
\raisebox{0.45\height}{\includegraphics[width=3.5cm]{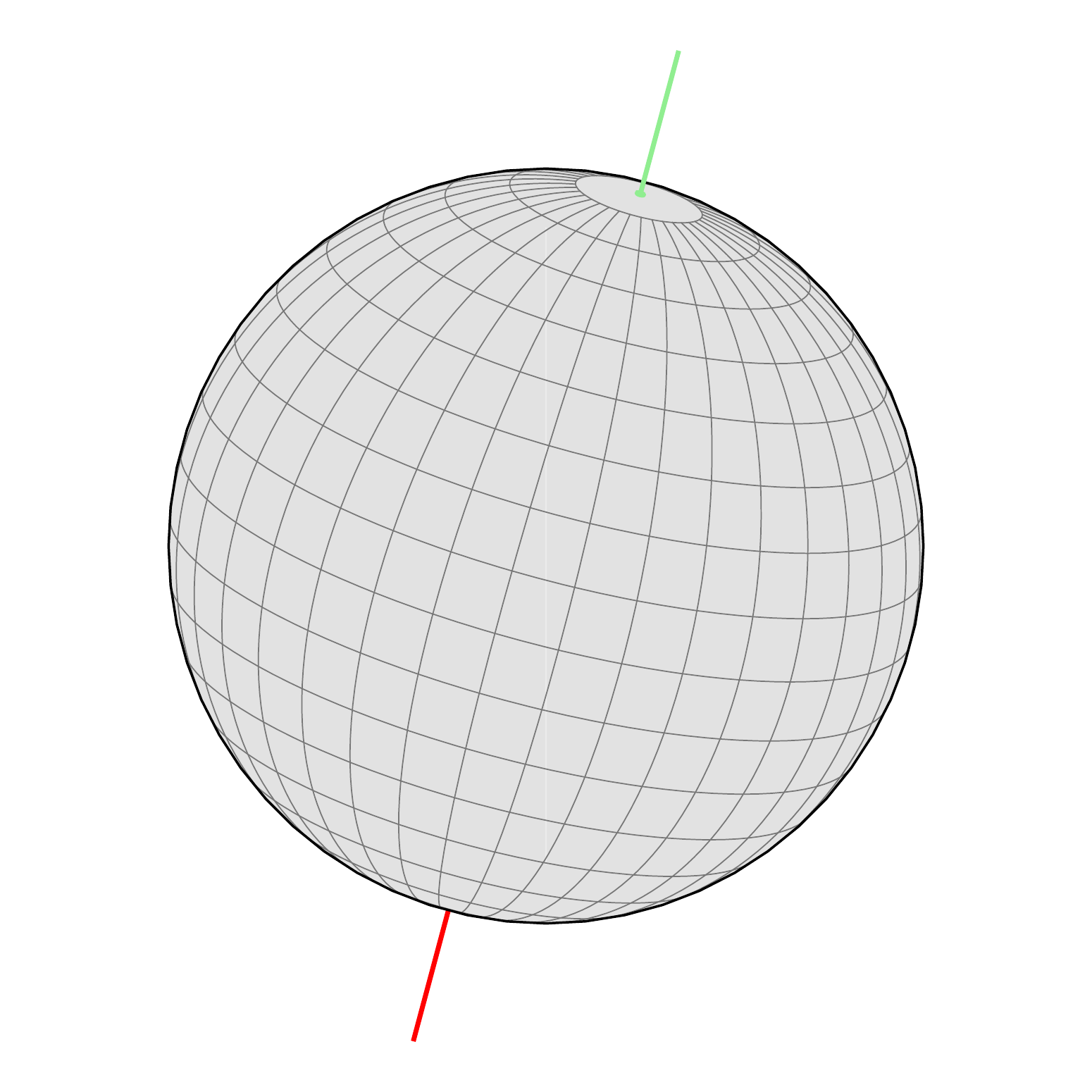}} \\
\end{tabular}
\caption{\emph{Left}: the best-fit oblique thermal models for \WASP{} (the free rotation case in Table \ref{table:fits}), in the 3.6 (top) and 4.5 (bottom) $\mu$m channels. The colored lines represent the best-fit light curve, with the shaded regions representing the 1- and 2-$\sigma$ uncertainty bounds (darker and lighter, respectively). The black points represent the binned data from \citet{cow12}. \emph{Center}: the density of explored axial orientations from the uncertainty calculations in the MCMC routine. The observer-facing hemisphere is framed by the dotted white lines. \emph{Right}: globes showing the best-fit orientation of the spin axes for each band, as viewed along our line of sight.}
\label{fig:WASP-12b_curves}
\end{center}
\end{figure*}

\begin{figure*}[htb!]
\begin{center}
\textbf{\COROT{}}\par\medskip
\begin{tabular}{rcc}
\includegraphics[width=7.5cm]{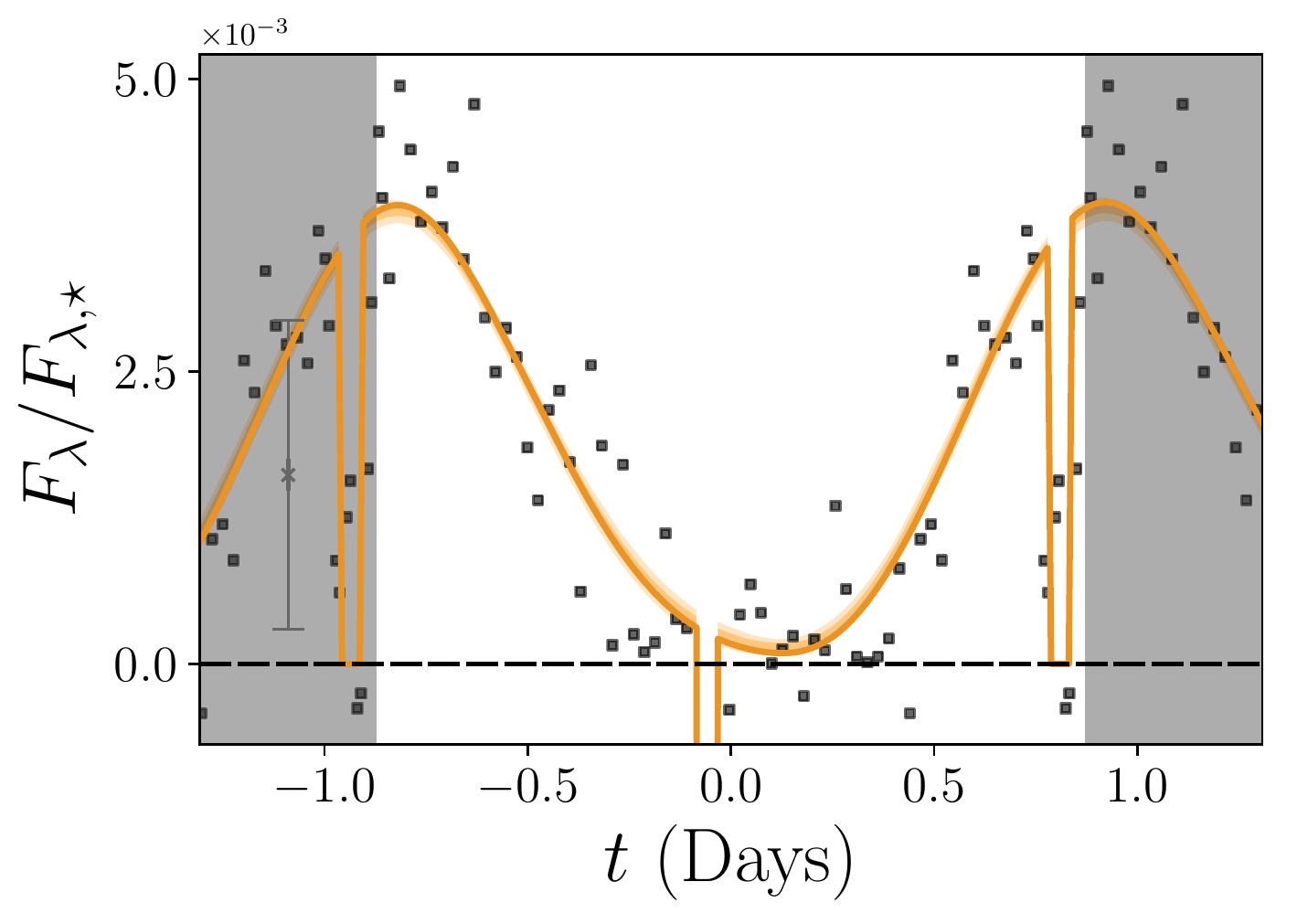} &
\raisebox{0.25\height}{\includegraphics[width=5.5cm]{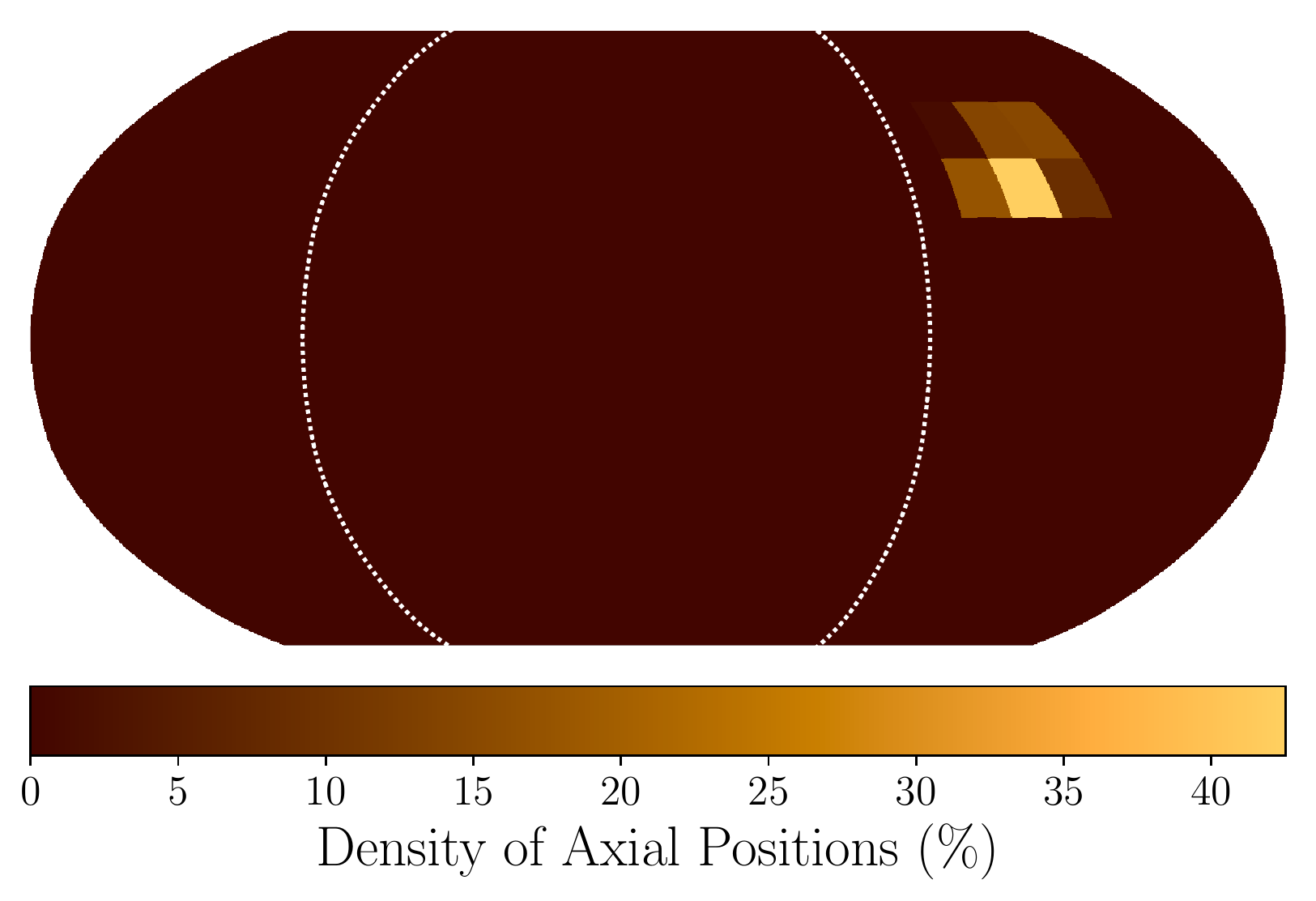}} &
\raisebox{0.45\height}{\includegraphics[width=3.5cm]{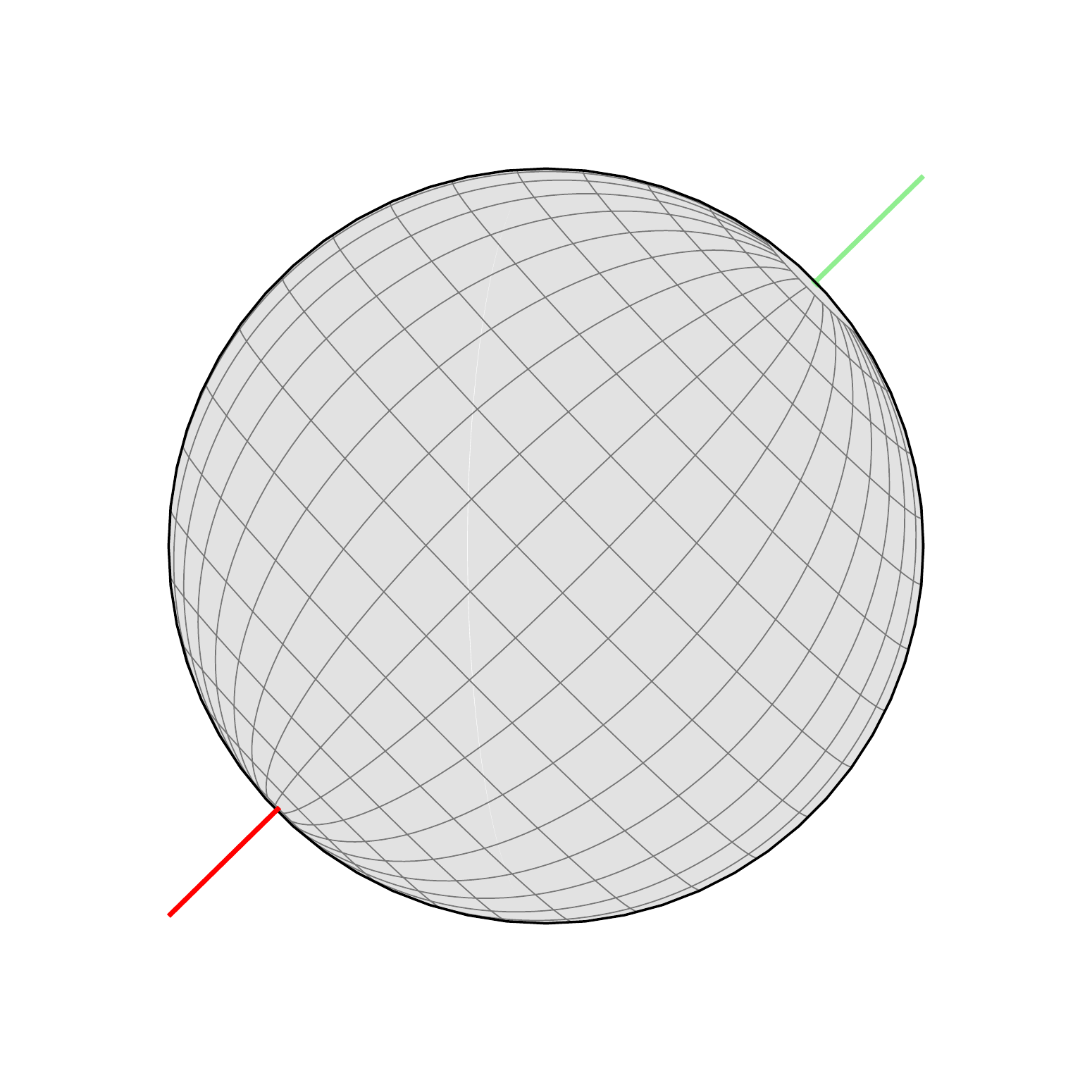}} \\
\end{tabular}
\caption{\emph{Left}: the best-fit oblique thermal model for \COROT{} (the sub-synchronous case in Table \ref{table:fits}), in the 4.5 $\mu$m channel. The colored lines represent the best-fit light curve, with the shaded regions representing the 1- and 2-$\sigma$ uncertainty bounds (darker and lighter, respectively). The black points represent the binned data from \citet{Dang2018}. \emph{Center}: the density of explored axial orientations from the uncertainty calculations in the MCMC routine. The observer-facing hemisphere is framed by the dotted white lines. \emph{Right}: a globe showing the best-fit orientation of the spin axis, as viewed along our line of sight.}
\label{fig:CoRoT-2b_curves}
\end{center}
\end{figure*}

\begin{deluxetable*}{ccccrrr}[htb!]
\tabletypesize{\footnotesize}
\tablewidth{0pt}
\tablecaption{Model Comparison by Number of Free Parameters}
\tablehead{\multicolumn{2}{c}{Obliquity} & \multicolumn{2}{c}{Rotation} & \multicolumn{3}{c}{$\Delta$AIC} \\ \colhead{Case} & \colhead{$\obliq$ ($^\circ$)} & \colhead{Case} & \colhead{$P_\mathrm{rot}/P_\mathrm{PSR}$} & \colhead{3.6 $\mu$m} & \colhead{4.5 $\mu$m} & \colhead{Combined}}
\startdata
\sidehead{\textbf{\HD{}}}
Fixed zero           & $0^\circ$ & Synchronous     & 1 &   2.62 &   0    &   1.37 \\
Fixed zero           & $0^\circ$ & Free            & $1.11\pm0.05$, $0.51^{+0.28}_{-0.03}$ &   0    &   1.26 &   0    \\
$\lambda$-consistent & $89.3^{+30.2}_{-31.8}$  & Sub-synchronous & 43.41 ($>1.20$) &   1.80 &   4.11 &   0.51 \\
$\lambda$-consistent & $119.0^{+41.5}_{-49.1}$ & Free            & $0.37^{+2.25}_{-0.04}$, $1.21^{+0.40}_{-0.68}$ &   3.51 &   6.32 &   4.39 \\
Free                 & $93.9^{+45.1}_{-32.5}$, $4.2^{+47.7}_{-1.9}$ & Sub-synchronous & $7.31$ ($>1.04$), $1.00^{+0.12}_{-0.00}$ &   1.78 &   4.12 &   4.63 \\
Free                 & $4.2^{+57.2}_{-2.7}$, $36.4^{+28.3}_{-3.4}$  & Free            & $1.09\pm0.06$, $0.39^{+0.08}_{-0.21}$ &   4.09 &   5.35 &   8.13 \\
\sidehead{\textbf{\WASP{}}}
Fixed zero           & $0^\circ$ & Synchronous     & 1 & 593.24 &  91.18 & 680.13 \\
Fixed zero           & $0^\circ$ & Free            & $0.91^{+0.01}_{-0.00}$, $0.95\pm0.01$ &   0    &   6.16 & 1.87 \\
$\lambda$-consistent & $19.4^{+1.7}_{-1.6}$  & Sub-synchronous & $1.002\pm0.001$ & 592.90 &  81.67 & 666.17 \\
$\lambda$-consistent & $11.1^{+4.5}_{-7.9}$  & Free            & $0.910^{+0.009}_{-0.001}$, $0.945^{+0.010}_{-0.005}$ &  68.37 &   6.63 &  66.57 \\
Free                 & $93.9^{+45.1}_{-32.5}$, $4.2^{+47.7}_{-1.9}$ & Sub-synchronous & $1.0015^{+0.0005}_{-0.0010}$, $1.0039^{+0.0011}_{-0.0009}$ & 586.68 &  76.60 & 658.98 \\
Free                 & $4.4^{+3.3}_{-2.2}$, $27.4^{+6.2}_{-10.8}$   & Free            & $0.91\pm0.004$, $0.95\pm0.01$ &   4.31 &   0    &   0    \\
\sidehead{\textbf{\COROT{}}}
Fixed zero           & $0^\circ$ & Synchronous     & 1 &        & 259.41 &        \\
Fixed zero           & $0^\circ$ & Free            & $1.10^{+0.02}_{-0.03}$ &        &   0    &        \\
Free                 & $45.8\pm1.4$        & Sub-synchronous & $1.07\pm0.01$ &        &   7.56 &        \\
Free                 & $2.0^{+8.1}_{-1.2}$ & Free            & $1.13^{+0.04}_{-0.02}$ &        &   4.06 &        \\
\enddata
\tablecomments{``$\lambda$-consistent'' refers to the models where the axial angles are consistent across each wavelength per MCMC step. ``Sub-synchronous'' indicates the rotation rate is entirely dependent on the obliquity angle, via Equation \ref{omega eq}. }
\label{table:AIC}
\end{deluxetable*}

For each planet we run two cases: one with 6 free parameters where the rotation rate is unconstrained, and another with 5 parameters where the rotation rate is fixed to the obliquity-dependent equilibrium rotation rate
\begin{equation}
\omega_{\mathrm{eq}} = n\frac{2\cos\epsilon}{1+\cos^2\epsilon}
\label{omega eq}
\end{equation}
where $n \equiv 2\pi/P_\mathrm{orb}$ is the mean motion (see \S\ref{Cassini state constraints} for more details). Additionally, for \HD{} and \WASP{}, which have data in two bands, we run the above two cases where we constrain the axial orientation to be consistent between the bands, for a total of 4 oblique model cases.

We use a Markov-Chain Monte Carlo process to broadly evaluate the likelihood landscape over the relevant parameters and converge on the sets of parameter values for each case with the most favorable likelihoods. Specifically, we employ a Metropolis-Hastings algorithm with simulated annealing \citep{Kirkpatrick1983}. Annealing introduces a temperature parameter that corresponds to acceptance probability, and serves to both broadly explore the likelihood space at initially high values, and to converge on optimal solutions as it is gradually reduced. After convergence, we continue the MCMC chain without annealing, to estimate uncertainties on the parameter values. The quoted 1-$\sigma$ uncertainties are determined by the 68\% ranges on either side of the values. The best-fit values and uncertainties for the models with band-distinct obliquities are listed in Table \ref{table:fits}. The resulting model light curves are plotted with these 1-$\sigma$ uncertainties, as well as the 2-$\sigma$ uncertainties, determined by the 95\% ranges, in the leftmost columns of Figures \ref{fig:HD149026b_curves}--\ref{fig:CoRoT-2b_curves}.

To evaluate the relative fit quality of these oblique models, we calculate the Akaike Information Criteria \citep{Akaike1973,Akaike1974} with a second-order correction for small sample sizes \citep{Sugiura1978,Hurvich1989,Hurvich1995}. This is given by
\begin{equation}
    \mathrm{AICc} \equiv 2 \left[k - \ln\mathcal{L} + \frac{k \left(k+1\right)}{n-k-1}\right],
\end{equation}
where $k$ is the number of free parameters, $\ln\mathcal{L}$ the log likelihood (and, by extension, the full second term $-2\ln\mathcal{L}$ being equal to the chi-squared statistic), and $n$ the sample size. We express the AIC values relative to the smallest value (i.e. the value of the most favorable model) in each band ($\Delta\mathrm{AICc} \equiv \mathrm{AICc} - \mathrm{AICc}_\mathrm{min}$), in Table \ref{table:AIC}. Additionally, to more appropriately compare the models with band-distinct obliquities with the models co-varying in obliquity, we calculate the cumulative AICc values for the combined model across both bands. Following the interpretation of \citet{Burnham2004}, $\Delta\mathrm{AICc}\leq2$ indicates a substantial level of evidence for a model (as compared with the most favorable); $4\leq\Delta\mathrm{AICc}\leq7$ implies low evidence, and $\Delta\mathrm{AICc}>10$ effectively implies no evidence. We discuss the results of each planet in the following sub-sections.

\subsection{\HD{}}\label{sec:results:HD149026b}
Following the prescription of the AIC, we find that the preferred model at 3.6 $\mu$m is the model with zero obliquity and unconstrained rotation. This implies that any fit improvement from adding axial orientation parameters are not statistically warranted. Indeed, the least-constrained model, which has a distinct obliquity and unconstrained rotation, returns a nearly upright axis, effectively approximating the non-oblique case. The $\Delta\mathrm{AIC}\sim 4$ is consistent with nearly identical likelihoods, since we are increasing the number of parameters by 2 by including obliquity.

In contrast, the most preferred model at 4.5 $\mu$m is the simplest --- zero obliquity and synchronous rotation. This implies that any discernible phase offset in the data is not strong enough to warrant a model that can capture it, either by a sub-synchronous rotation or high obliquity. The non-oblique, unconstrained rotation case is next in line, with a best-fit spin-orbit ratio of nearly two. Moving into the oblique cases, we see that there is a moderate drop-off in support for the models at 4.5 $\mu$m; this is not surprising, since the light curves at 3.6 and 4.5 $\mu$m disagree with the direction of offset.

Interestingly, when we compare the combined model criteria, the combined non-oblique model with unconstrained rotation is on top, but the model with consistent obliquity and sub-synchronous rotation is not far behind. The most likely interpretation is that the 3.6 $\mu$m data are modestly better fit by an axis nearly perpendicular to the orbit normal, rather than simply a slow effective rotation rate. Given the very slow returned rotation for this case, it also suggests that the rotation has a weak effect on the offset at this high of obliquity, since the transverse motion is taken almost completely out of the plane of the sky from our perspective.

\subsection{\WASP{}}\label{sec:results:WASP-12b}
Due to its strong tidal distortion, \WASP{} is the most complicated of the planets, and we expect that it will be difficult to fully capture its shape only accounting for rotation and obliquity. In sharp contrast with the moderate differences in $\Delta\mathrm{AIC}$ between models in the bands of \HD{}, which imply a dominant effect of the number of parameters, for \WASP{} both the synchronous and sub-synchronous models are vastly worse than those with unconstrained rotation. This makes sense given that the light curves in both bands show strong variations between occultations, at least one of which can be approximated using an eastward phase offset. Eastward offsets are more amenable to a super-rotating atmospheric layer than high obliquity, and indeed the models with the most free parameters return small values for obliquity.

AIC however cannot provide a quantification of the absolute fit quality. While our model light curves generally lie within the range of observed fluxes at any given time, the maximum observed amplitude of semi-annual variations exceed what can be captured with a combination of spin-orbit geometry and thermal modeling. Ellipsoidal variations due to the planet's tidal distortion should only change the observed surface area by $\sim\!10\%$ \citep{lis10,lai10}, and the complementary effects on the star's shape should be a further order of magnitude smaller. These correspond to variations on the orders of $\sim\!4\times10^{-4}$ and $\sim\!4\times10^{-5}$, according to \citet{cow12}. Indeed, the incorporation of an ellipsoidal distortion to the non-oblique thermal model \citep[e.g.~in][]{ada18b} does not provide enough additional variation to approach the full amplitudes of the data.

\subsection{\COROT{}}\label{sec:results:CoRoT-2b}
Our analysis of the light curves is more limited for \COROT{} since it only has one band, but the AIC values do indicate that the non-synchronous models are much better fits to the data. As with the 3.6 $\mu$m fits for \HD{}, the differences in selection criteria between the non-oblique, non-synchronous and oblique models appear to be largely influenced by the AIC's penalty for additional parameters by introducing axial orientation. From this we infer that the model fits from either a westward wind interpretation (slower than synchronous rotation with little to no obliquity) or a sub-synchronous rotation at moderately high obliquity are of similarly effective quality.

Considering these three cases, we can draw some conclusions. For planets whose phase variations are already amenable to fitting by a non-oblique thermal model, but whose observed westward phase offsets require a slower-than-synchronous effective rotation rate (most readily interpreted as westward winds), our fit quality is at least as good with an oblique model where we assume an obliquity-dependent sub-synchronous rotation (Equation \ref{omega eq}). However, particularly for \WASP{}, we have no cases where the addition of obliquity improves the quality of fit of a thermal model which has difficulty reproducing all major features of the data.

\section{Dynamical Evolution of a Tightly Packed System}\label{sec:REBOUND}
Having established the feasibility of obliquity to fit the observed phase variations, we now place this interpretation in a dynamical framework in the following sections. \citet{Batygin2016} set a physical framework where a system comprising multiple close-in planets (i.e. $P \lesssim 100$ days) with masses in the super-Earth regime ($1 < M/M_\oplus < 30$) can evolve to have high mutual orbital inclinations. They first determine that, above an initial mass of $\gtrsim 15 M_\oplus$, the innermost planet can undergo runaway accretion under a range of stellar nebular densities, reaching $M_p \sim M_{\mathrm{J}}$ within a timescale $\sim\!10^6$ years, thereby predicting in-situ formation of close-in giant planets. Their model couples this with the evolution of the host star onto the main sequence and determines that outer planets can evolve to high mutual inclinations via nodal regression commensurability. These results constitute a prediction that systems with hot Jupiters and mutually inclined outer planets should be common.

In order to evaluate the feasibility of evolving a dynamically full multi-planet system into a configuration consistent with the existence of secular spin-orbit resonance and associated high obliquities, we performed dynamical simulations using the \texttt{REBOUND}\footnote{\url{https://github.com/hannorein/rebound}} software package with the \texttt{IAS15} integrator \citep{rei12,rei15}. For our fiducial precursor system, we adopted a known multi-planet system, Kepler-107 \citep{Rowe2014,VanEylen2015}. We then statistically explored the dynamical reaction of the configuration to orbital instability triggered by rapid increase in the mass of the innermost planet. We assumed an instantaneous increase from the estimated innermost planet mass of 3.7 $M_\oplus$, to that of Jupiter. We also integrated the system without any mass increase. For both cases we performed 100 separate simulations with a duration of at least $\sim\!10^5$ Earth years ($\sim\!10^7$ orbits of the innermost planet as initialized).

\begin{deluxetable*}{cccccccc}
\tabletypesize{\footnotesize}
\tablewidth{0pt}
\tablecaption{Orbital and Transit Properties for Kepler-107}
\tablehead{\colhead{Planet} & \colhead{$P$ (days)} & \colhead{$a$ (AU)} & \colhead{$e$} & \colhead{$M_\mathrm{p}$ ($M_J$)} & \colhead{$R_\mathrm{p}$ ($R_J$)} & \colhead{$b$} & \colhead{$T_\mathrm{tran}$ ($\mathrm{BJD}-2454900$)}}
\startdata
b & $3.179997\pm1.1\times10^{-5}$ & 0.044 & $0.020^{+0.200}_{-0.020}$ & 0.01167 & $0.139\pm0.005$ & $0.34\pm0.24$ & $66.49904\pm0.00200$ \\
c & $4.901425\pm1.6\times10^{-5}$ & 0.059 & $0.020^{+0.260}_{-0.020}$ & 0.0133 & $0.161\pm0.016$ & $0.78\pm0.29$ & $71.60742\pm0.00174$ \\
d & $7.958203\pm1.04\times10^{-4}$ & 0.082 & $0.14^{+0.25}_{-0.14}$ & 0.00371 & $0.095\pm0.005$ & $0.27\pm0.24$ & $70.79968\pm0.00612$ \\
e & $14.749049\pm3.4\times10^{-5}$ & 0.123 & $0.020^{+0.180}_{-0.020}$ & 0.0360 & $0.308\pm0.022$ & $0.90\pm0.28$ & $71.77998\pm0.00134$ \\
\enddata
\tablerefs{Orbital eccentricities are taken from \citet{VanEylen2015}. The masses are calculated from an empirical relation calculated on \url{exoplanets.org}, detailed in \citet{hane14}. All other values are from \citet{Rowe2014}.}
\label{table:Kepler-107}
\end{deluxetable*}

\begin{figure*}[htb!]
\begin{center}
\begin{tabular}{c}
\includegraphics[width=17cm]{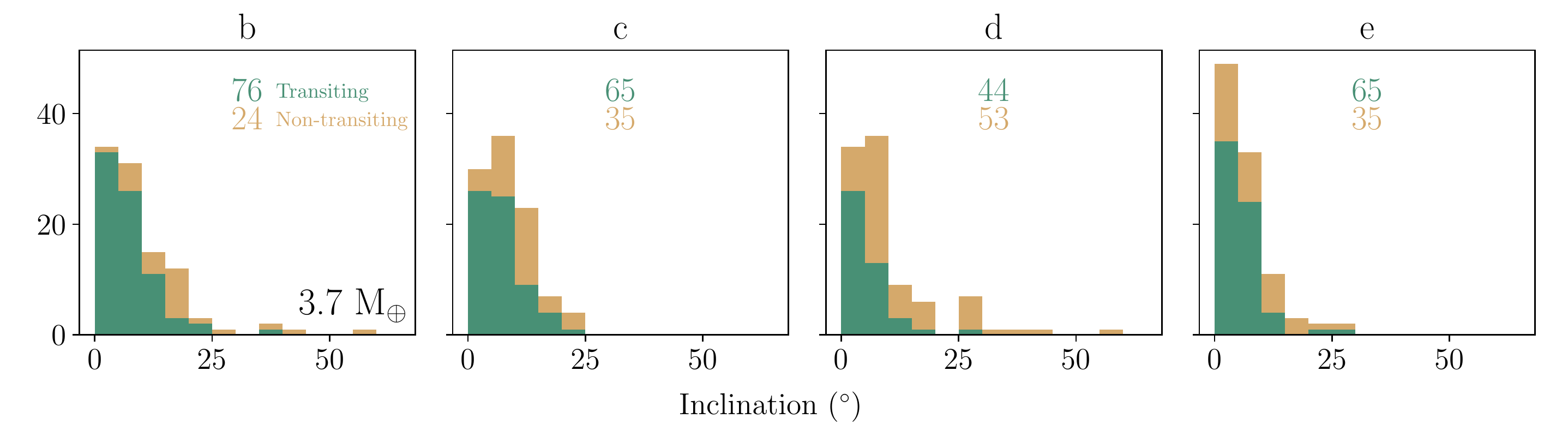} \\
\includegraphics[width=17cm]{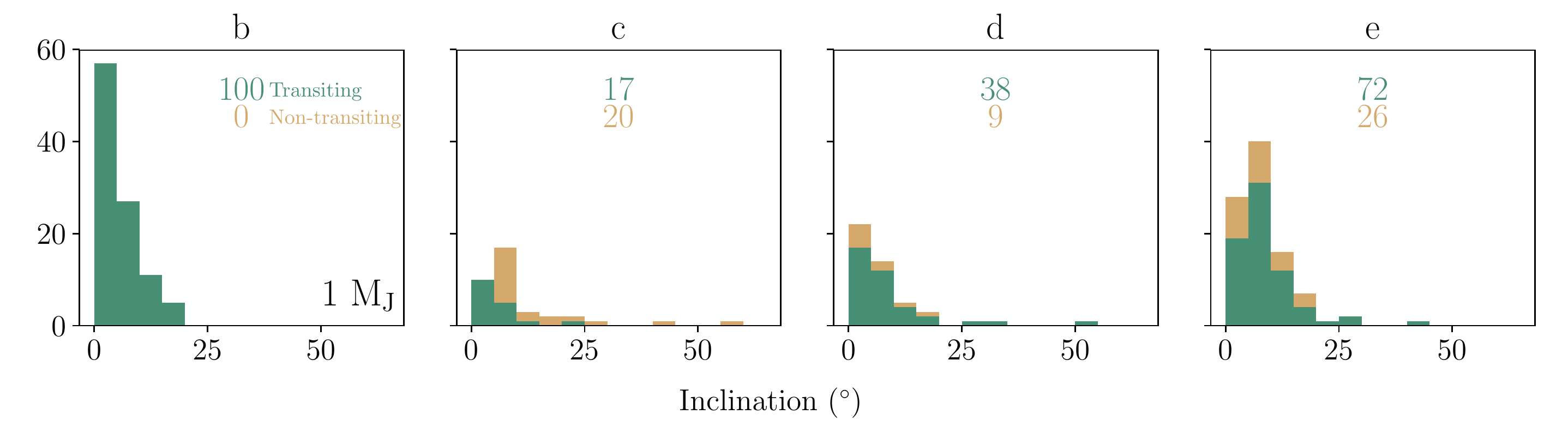} \\
\end{tabular}
\caption{\emph{Top}: The distribution of mutual inclinations (in degrees) from the net angular momentum vector for the nominal Kepler-107 system randomized 100 times over orbital parameters consistent with the observed transit parameters. \emph{Bottom}: The same distributions where the mass of planet $b$ is increased from its estimated 3.7 M$_\oplus$ to that of Jupiter. Each simulation was run for $\sim\!10^5$ Earth years, corresponding to $\sim\!10^7$ orbital periods (as initialized) of the innermost planet. Green entries denote planets which still transit according to all the calculated orbital parameters; goldenrod entries denote those that will not transit. The total number of simulations in each box may not sum to 100, due to ejections.}
\label{fig:inclination_dists}
\end{center}
\end{figure*}

The orbital geometry of the Kepler-107 system is not entirely constrained by the transit parameters reported by \citet{Rowe2014}. We therefore adopted initial values for the semi-major axes from \citet{Rowe2014} and initial eccentricities from \citet{VanEylen2015}. For the angular orbital elements, we used the known transit parameters in conjunction with the following process to randomly generate orbital elements consistent with the observational constraints.
\begin{enumerate}
\item Set the line of sight to the positive $x$-axis, the default reference direction in \texttt{REBOUND}\footnote{Note that in this setup, an inclination of zero will imply the orbit is viewed edge-on, rather than the standard $i=90^\circ$. We adopt this non-standard definition purely for convenience.}.
\item Choose a longitude of periastron, $\varpi \sim U\!\left(0,2\pi\right)$.
\item Choose a longitude of ascending node $\Omega$ from a uniform distribution over the set of angles satisfying
\begin{equation}
\lvert\sin\Omega\rvert \geq b \left[\frac{R_\star+R_\mathrm{P}}{r_\mathrm{tr}\!\left(e, \varpi\right)}\right]
\end{equation}
where $b$ is the impact parameter, $R_\star$ and $R_\mathrm{P}$ the stellar and planetary radii, respectively, and
\begin{equation}
r_\mathrm{tr} \equiv a\left(\frac{1-e^2}{1+e\sin\varpi}\right)
\end{equation}
the star-planet separation at the transit mid-point.
\item Calculate the inclination in the range $\left[0, \pi/2\right]$ that solves the transit condition
\begin{equation}
\sin i \, \lvert\sin\Omega\rvert = b \left[\frac{R_\star+R_\mathrm{P}}{r_\mathrm{tr}\!\left(e, \varpi\right)}\right].
\end{equation}
\end{enumerate}

After integrating systems constructed in this manner for $\sim\!10^5$ Earth years, we see a significant evolution of the systems away from their initially nearly co-planar states. We calculate the inclinations (relative to the net system angular momentum vector) for the final orbits of the remaining planets and evaluate the transit conditions along the initial line of sight (Figure \ref{fig:inclination_dists}). In a significant fraction of simulations, some combination of the outer planets ($c$--$e$) no longer transit due to large inclinations which can exceed 20$^\circ$. Out of the 100 simulations with the nominal Kepler-107 masses, only 3 saw an ejection (defined in our simulation is defined as reaching a separation from the star an order of magnitude larger than the initial semi-major axis). The fraction of ejections increases greatly in the inner-Jupiter model, with a significant fraction of trials ejecting $c$ and/or $d$ within $10^5$ years. Half of ejections occurred within the first $\sim\!2.5\times10^4$ years. These results as a whole suggest that, under a significant perturbation such the runaway accretion of the inner planet, it is plausible that a compact multi-transiting planet system such as Kepler-107 could evolve to high mutual inclinations and/or ejections for the outer planets, leaving a majority of cases where only the inner planet could be observed to transit.

The simulations presented here have only accounted for the radii of the star and orbiting planets in terms of collision detection, whose prescribed resolution was a merger. No mergers occurred in any trial. Otherwise, the bodies are effectively treated as point masses. If these bodies are endowed with structure, as formulated for example by \citet{mar02}, then secular inclination resonances can act to increase the degree of dynamical instability (leading to agglomerating collisions) and mutual inclinations among remaining planets, thereby increasing the fractions of non-transiting planets \citep{Batygin2016}. \\

\section{Cassini State Driven by an Inclined, External Perturber}\label{sec:Cassini state}

While we have examined the possibility that close-in giant planets have non-zero obliquities, we have not yet discussed in detail how such states may arise. Tidal torques are strongest for these close-in, large-radius planets, and they act to dampen planetary obliquities to zero. Large tilts can be maintained, however, if a planet is locked in a secular spin-orbit resonance involving synchronous precession of its spin vector and orbital angular momentum vector \citep{Fabrycky2007}. Of particular interest is Cassini State 2 \citep{Peale1969}, where the planet maintains a large obliquity as the spin and orbital axes precess at the same rate on opposite sides of the axis normal to the invariable plane. In the absence of dissipation, these three axes are coplanar, but in the presence of tides, the spin axis is slightly shifted out of the plane.

In this section, we evaluate the possibility that the planets \HD{}, \WASP{}, and \COROT{}  have their spin axes trapped in Cassini states due to spin-orbit resonances driven by exterior perturbing planets. The case of a resonance for \WASP{} was already discussed in detail by \cite{MillhollandLaughlin2018} as a theory for the planet's rapid orbital decay \citep{mac2016, Patra2017, Maciejewski2018, Bailey2019}. We include elements of \cite{MillhollandLaughlin2018}'s analysis here for consistency and comparison with the other two systems. 

We start by calculating the frequencies of spin and orbital precession. We then examine the constraints on the hypothetical perturbing planets in order for the spin-orbit resonant configurations to be plausible. Table \ref{system parameter table} shows the system parameters we adopt in these calculations.

\subsection{Spin-Orbit Resonant Frequencies}

The torque from the host star on a rotationally-flattened planet will cause the planet's spin-axis to precess about the orbit normal at a period, $T_{\alpha} = 2\pi/(\alpha\cos\epsilon)$. Here $\epsilon$ is the obliquity and $\alpha$ is the precession constant \citep{Ward2004}, which is given by  

\begin{equation}
\alpha = \frac{1}{2}\frac{M_{\star}}{M_{\mathrm{p}}}\left(\frac{R_{\mathrm{p}}}{a}\right)^3\frac{k_2}{C}\omega.
\label{alpha}
\end{equation}
We have assumed there are no satellites and have defined $M_{\mathrm{p}}$, the planet mass, $C$, the moment of inertia normalized by $M_{\mathrm{p}} {R_{\mathrm{p}}}^2$, and $\omega$, the spin frequency. This expression also assumes that the coefficient of the quadrupole moment of the planet's gravitational field, $J_2$, takes the form \citep{Ragozzine2009},
\begin{equation}
J_2 = \frac{\omega^2 {R_{\mathrm{p}}}^3}{3 G {M_{\mathrm{p}}}}k_2.
\end{equation}

The secular spin-orbit resonance requires a commensurability between the planet's spin-axis precession frequency and its orbit nodal regression frequency, $g = \dot{\Omega}$. Planet-planet interactions are one source of nodal regression. In a two-planet system, the nodes of both planets regress uniformly due to secular perturbations. The frequency is given by Laplace-Lagrange theory to be
\begin{align}\label{LL g}
\begin{split}
g_{_\mathrm{LL}} = &-\frac{1}{4} b_{3/2}^{(1)}(\alpha_{12})\alpha_{12} \times \\
&\left(n_1\frac{M_{p2}}{M_{\star} + M_{p1}}\alpha_{12} + n_2\frac{M_{p1}}{M_{\star} + M_{p2}}\right),
\end{split}
\end{align} 
if the planets are not near mean-motion resonance \citep{1999ssd..book.....M}. Here, $\alpha_{12} = a_1/a_2$ and $n_i$ is the mean-motion of planet $i$, $n_i^2 = G M_{\star}/a_i^3$. The constant, $b_{3/2}^{(1)}(\alpha_{12})$ is a Laplace coefficient, defined by 
\begin{equation}
b_{3/2}^{(1)}(\alpha_{12}) = \frac{1}{\pi}\int_{0}^{2\pi}\frac{\cos\psi}{(1-2\alpha\cos\psi + \alpha^2)^{3/2}}d\psi.
\end{equation}

In addition to planet-planet interactions, the stellar quadrupole gravitational moment also induces orbit nodal recession about the stellar spin vector. In the absence of secular planet interactions, this occurs at the frequency \citep{Spalding2017}
\begin{equation}
g_{\star} = n \frac{k_{2\star}}{2}\left(\frac{\omega_{\star}}{n}\right)^2\left(\frac{R_{\star}}{a}\right)^5.
\label{star g}
\end{equation} 
Accordingly, in a multiple-planet system, the planets' orbit normal vectors evolve in response to perturbation components at different frequencies. One frequency, $g_{p-p}$, is due to the planet-planet perturbations and the other, $g_{\star}$, is associated with the stellar quadrupole moment. These are close but not exactly equal to the equations \ref{LL g} and \ref{star g} above, since the analytical expressions only account for one driver of nodal recession at a time. When the stellar equatorial plane is coincident with the plane perpendicular to the total orbital angular momentum vector, these two frequencies add linearly such that $g= g_{p-p} + g_{\star}$, and the nodal recession is uniform. However, if these planes are not coincident (i.e. if there is an angle between the stellar spin vector and the total orbital angular momentum vector), then the frequencies do not add linearly, but rather the nodal recession is a non-uniform superposition of these modes.

The spin-orbit resonance can be encountered and captured when the spin precession is commensurable with either one of these orbital frequency components, $g_{p-p}$ or $g_{\star}$  \citep{MillhollandLaughlin2019}. In the case of short-period planets, however, the spin axis precession is so fast that spin-orbit resonances induced by planet-planet interactions are much more likely. Although $g_{\star}$ is fast early in the system's lifetime because the star is rapidly rotating and has not finished contracting \citep{Batygin2013, Spalding2017}, $g_{\star}$ is significantly smaller for a main-sequence star, and it is generally not fast enough for spin-orbit resonance. To illustrate this explicitly, we calculate $g_{\star}$ for our three case studies. The results are shown in Table \ref{system parameter table}. In all three systems, $g_{\star}$ is much smaller than the spin axis precession constant. This clearly indicates that nodal recession must be provided by planet-planet perturbations if these systems are in spin-orbit resonances. We will therefore assume in the remainder of this work that planet-planet resonances are the relevant ones. \\

\subsection{Constraints on the masses and semi-major axes of potential inclined companions}

Suppose that, in each of the three systems, planet $b$ is in a high-obliquity spin-orbit resonance with its orbital precession induced by secular interactions with an exterior, inclined, and as-yet undetected companion. Existence of this configuration places significant constraints on the masses, semi-major axes, and inclinations of the perturbing companions. These constraints stem from \begin{enumerate*}[label={(\roman*)}]
\item upholding the Cassini state, \item preserving total angular momentum conservation, and \item maintaining consistency with existing radial velocity (RV) data.\end{enumerate*} In the subsections that follow, we review and apply each of these constraints in detail. 

\subsubsection{Constraints from Maintaining the Cassini State}
\label{Cassini state constraints}

\begin{deluxetable}{cccc}
\tabletypesize{\footnotesize}
\tablewidth{0pt}
\tablecaption{System parameters used in developing the Figure \ref{Obliquity heatmap} constraints on the perturbing planet parameters. The quantities \textbf{$k_{2\star}$}, $k_2$, and $C$ are estimates. $\alpha_{\mathrm{syn}}$ is the spin axis precession constant that would result from the aggregate of system parameters in the case of synchronous rotation ($\omega = \alpha_{\mathrm{syn}}$). $g_{\star}$ is the approximate frequency of nodal recession induced by stellar oblateness. Note that $\alpha_{\mathrm{syn}} \gg g_{\star}$ in all cases.}
\tablehead{\colhead{} & \colhead{\HD{}} & \colhead{\WASP{}} & \colhead{\COROT{}}}
\startdata
$M_{\star} \ [M_{\odot}]$ &  $1.345^{(1)}$ & $1.36^{(2)}$ & $0.97^{(3)}$ \\
$R_{\star} \ [R_{\odot}]$ &  $1.541^{(1)}$ & $1.63^{(2)}$ & $0.90^{(3)}$ \\
$P_{\star} \ [\mathrm{days}]$ &  $\sim 13^{(4^*)}$ & $36^{(5)}$ & $4.5^{(3)}$ \\
$k_{2\star}$ & 0.01$^{(6)}$ & 0.01$^{(6)}$ & 0.01$^{(6)}$ \\
$a \ [\mathrm{AU}]$ & $0.042^{(4)}$ & $0.02299^{(7)}$ & $0.0281^{(3)}$ \\
$M_{\rm p} \ [M_{\rm {Jup}}]$ &  $0.36^{(4)}$  & $1.41^{(7)}$ & $3.3^{(3)}$ \\
$R_{\rm p} \ [R_{\rm {Jup}}]$ &  $0.725^{(4)}$ & $1.89^{(2)}$ & $1.466^{(3)}$ \\
$k_2$ & 0.1 & 0.1 & 0.1 \\
$C$ & 0.2 & 0.2 & 0.2 \\
\hline
$2\pi/{\alpha_{\mathrm{syn}}} \ \mathrm{[yr]}$ & 14.3 & 0.2 & 4.0\\
$2\pi/{g_{\star}} \ \mathrm{[yr]}$ & $2.4\times10^5$ & $1.7\times10^5$ & $8.7\times10^4$\\
\enddata
\tablerefs{$^{(1)}$\cite{Carter2009}
$^{(2)}$\cite{Maciejewski2013}
$^{(3)}$\cite{Bonomo2017}
$^{(4)}$\cite{Sato2005} 
$^{(5)}$\cite{Watson2010}
$^{(6)}$Estimate using \cite{Batygin2013}
$^{(7)}$\cite{heb09} 
$^{(*)}$Estimate using $v\sin i$}

\label{system parameter table}
\end{deluxetable}

If a planet is captured in a Cassini state, there will be a resonant commensurability that can be stated as \citep{Ward2004}
\begin{equation}
\lvert g \rvert \approx \alpha\cos\epsilon.
\label{resonance condition}
\end{equation}
This holds if the planetary obliquity, $\epsilon$, is large compared to the mutual inclination between the orbits. We will take that as an assumption and use this condition to calculate the range of values of each perturbing planet's mass, $M_{p2}$, and semi-major axis, $a_2$, that allow for resonant commensurability. Table \ref{system parameter table} shows the estimates of the spin axis precession periods used in these calculations.

\begin{figure}
\epsscale{1.25}
\plotone{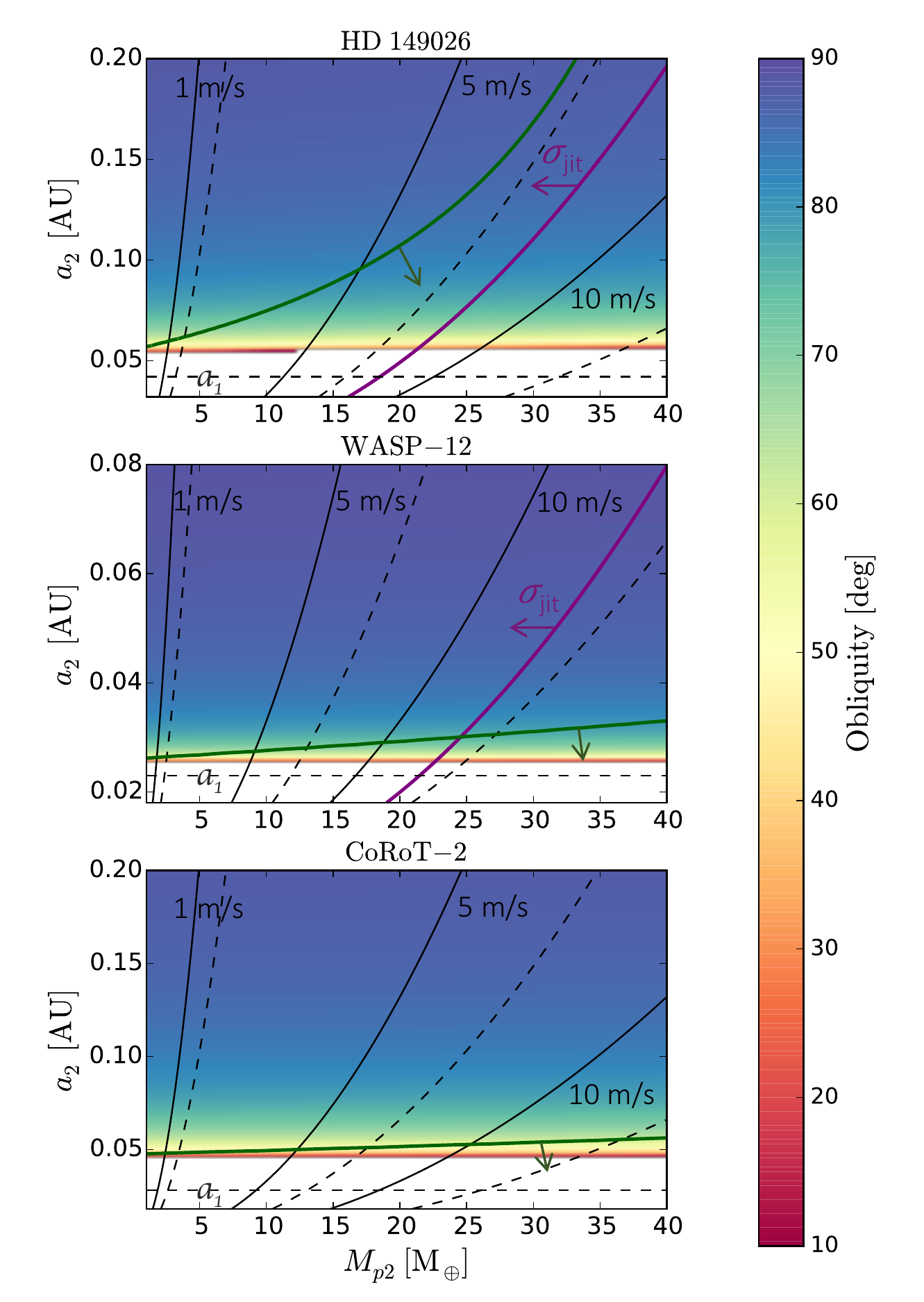}
\caption{A map of the obliquities that \HD{}, \WASP{}, and \COROT{} would require if they were captured in a secular spin-orbit resonance with an external perturber of mass, $M_{p2}$, and semi-major axis, $a_2$. The black solid/dashed contours represent the companion planet's RV semi-amplitude for $i_2=90^{\circ}$/$i_2=45^{\circ}$, respectively. The solid purple lines are contours of $\sigma_{\mathrm{jit}}$, the RV jitter of the fit to planet b, which yields the approximate detection limit. (The $\sigma_{\mathrm{jit}}$ contour for \COROT{} is so large that it is off the plot.) The green lines are conservative constraints arising from the assumption that angular momentum conservation is entirely maintained by damping the mutual inclination of the two planets' orbit planes (see section \ref{angular momentum conservation}). Note that the second panel is similar to Figure 3 of \cite{MillhollandLaughlin2018}.} 
\label{Obliquity heatmap}
\end{figure}

In Figure \ref{Obliquity heatmap}, we show heatmaps in $M_{p2}$ and $a_2$ space that represent the obliquity of planet $b$ necessary for the resonance to hold in each system. We assume that the nodal recession is driven by interactions with the secondary planet, such that $g = g_{p-p} \approx g_{_\mathrm{LL}}$. We also assume $e_b = 0$ and that the spin rate of planet $b$ is at equilibrium, at which $d\omega/dt = 0$. The equilibrium rate is given by equation \ref{omega eq} in the traditional viscous approach to equilibrium tide theory \citep{Levrard2007}. The quantity $n\equiv2\pi/P_\mathrm{orb}$ is the mean-motion.
Combining equations \ref{alpha}, \ref{resonance condition}, and \ref{omega eq}, the resonance condition becomes 
\begin{equation}
\lvert g \rvert = \alpha_{\mathrm{syn}}\frac{2\cos^2\epsilon}{1+\cos^2\epsilon},
\end{equation}
where $\alpha_{\mathrm{syn}}$ is the value of $\alpha$ in the case of synchronous rotation, $\omega = n$. The solution for $\epsilon$ is then 
\begin{equation}
\cos\epsilon = \left(\frac{1}{2\alpha_{\mathrm{syn}}/{\lvert g \rvert} - 1}\right)^{1/2}.
\end{equation}
This expression can be used to calculate $\epsilon$ for a range of values of $M_{p2}$ and $a_2$.

Figure \ref{Obliquity heatmap} shows that if $a_2$ is too small, $\lvert g \vert$ is too large and no resonance is possible (white regions). Alternatively, if $a_2$ is too large, $\lvert g \vert$ is small and $\epsilon \sim 90^{\circ}$, which is unstable in the long-term. Figure \ref{Obliquity heatmap} thus allows us to define approximate constraints on the semi-major axes of the hypothetical companion planets. The constraints on their masses are not strong due to the weak dependence of $g_{_\mathrm{LL}}$ on $M_{p2}$ when $M_{p1} \gg M_{p2}$.

\subsubsection{Constraints from Angular Momentum Conservation}\label{angular momentum conservation}

For the equilibrium tidal theory that we are considering, a planet maintained in an oblique state spins sub-synchronously (equation \ref{omega eq}), and the rate at which tides convert orbital energy into heat energy is orders of magnitude larger than it would be in the case of zero obliquity \citep{Levrard2007, Wisdom2008}. An oblique tidally dissipating planet migrates inwards towards the star, decreasing $\lvert g \rvert/\alpha$ in the process and further increasing its obliquity. The decrease in orbital angular momentum associated with inward migration must be counteracted such that total angular momentum of the system, $\mathbf{J}$, is conserved. This can be accomplished through gradual alignment of the planetary orbital angular momenta, $\mathbf{L_1}$ and $\mathbf{L_2}$, with one another and with the stellar spin angular momentum, $\mathbf{S_{\star}}$. If we assume for a conservative argument that alignment with $\mathbf{S_{\star}}$ does not play a role, then the second planet must have enough angular momentum to preserve the total conservation. This allows conservative limits to be placed on the orbits of the perturbing planets \citep{Fabrycky2007}.

We begin with an expression for $\mathbf{J}$. We define $i$ to be the angle between $\mathbf{L_1}$ and $\mathbf{L_2}$. In addition, we define $\phi$ to be the angle between $\mathbf{S_{\star}}$ and $\mathbf{L_1}+\mathbf{L_2}$. Then, assuming the planets' own spin angular momenta are negligible, the magnitude of $\mathbf{J}$ is given by 
\begin{equation}
\begin{split}
\label{Jsq}
J^2 &= {S_{\star}}^2 + {L_1}^2 + {L_2}^2 + 2{L_1}{L_2}\cos i \\
&+ 2{S_{\star}}({L_1}^2 + {L_2}^2 + 2{L_1}{L_2}\cos i)^{1/2}\cos\phi.
\end{split}
\end{equation}

Secular interactions between the planets do not change $a_2$ or $e_2$ to first order, so $L_2$ remains fixed. Therefore, as $L_1$ decreases, conservation of $J$ must be upheld by an increase in $S_{\star}$ via tidal spin-up of the star \citep{Brown2011} or a decrease in $i$ and $\phi$ through reorientation of the orbits. As stated above, we will conservatively assume that stellar spin-up and reorientation between the orbital and stellar spin angular momenta do not play a role. Under this assumption, the maximum possible value of planet $b$'s initial semi-major axis, $\max({a_{1i}})$, may be expressed in terms of $a_2$ and $M_{p2}$, 
\begin{equation}
\max({a_{1i}}) = a_1\left[1 + 2\left(\frac{M_{p2}}{M_{p1}}\right)\left(\frac{a_2}{a_1}\right)^{1/2}\right],
\label{max a_1i}
\end{equation}
where $a_1$ corresponds to the present-day value of the semi-major axis. This expression is obtained by assuming that $i$ was initially near $90^{\circ}$ and is currently near $0^{\circ}$. We also assumed $e_2 \approx 0$.

This expression does not hold enough information to constrain $a_2$ and $M_{p2}$ in and of itself, since there is no clear limit on $\max(a_1)$. To develop a constraint, we apply the additional requirement that the initial obliquity of planet $b$ must be near $0^{\circ}$. Though this is not strictly necessary, if it is true, it makes the initial resonant capture scenario easy to explain. (Recall that $\epsilon$ increases as the planet tidally migrates inwards.) For a given $a_2$ and $M_{p2}$, the semi-major axis, $a_{1i}$, at which the initial obliquity is zero satisfies the expression
\begin{equation}
g(a_{1i},a_2) = \alpha_1(a_{1i}).
\end{equation} 
There is only a plausible solution if $a_{1i} <= \max({a_{1i}})$. Accordingly, this is the additional constraint that we use in conjunction with equation \ref{max a_1i}.

The green lines in Figure \ref{Obliquity heatmap} delineate the region that simultaneously upholds the conservative angular momentum limits and maintains the possibility of a small initial obliquity for planet $b$. Significant regions of phase space are ruled out for both the WASP-12 and CoRoT-2 systems. While this is somewhat problematic for the theory, it is important to keep in mind that these tight constraints can be strongly alleviated and/or removed by assuming a non-zero initial obliquity or by allowing realignment with $\mathbf{S_{\star}}$ to account for some degree of the system's angular momentum conservation.

\subsubsection{Constraints from Radial Velocity Data}

An entirely unrelated constraint arises from the set of RV measurements of the systems. The hypothetical companion planets must have RV semi-amplitudes smaller than the current detection limits; otherwise they would have already been discovered. Table \ref{RV jitter table} shows current estimates of the RV semi-amplitudes, $K_1$, of the known planets and the RV jitters of the fits, $\sigma_{\mathrm{jit}}$. The hypothetical outer planets must have $K_2 \lesssim \sigma_{\mathrm{jit}}$. Figure \ref{Obliquity heatmap} shows that the parameter space in the HD 149026 and WASP-12 systems are somewhat constrained by the RV limits. CoRoT-2's jitter, however, is so large that hardly any phase space area is ruled out. 

\begin{deluxetable}{cccc}
\tabletypesize{\footnotesize}
\tablewidth{0pt}
\tablecaption{RV semi-amplitudes of the known hot Jupiters in each system, along with RV jitters of the fits.}
\tablehead{\colhead{} & \colhead{$K_1$ [m/s]} & \colhead{$\sigma_{\mathrm{jit}}$ [m/s]} & \colhead{Reference}}
\startdata
\HD{} & $38.5 \pm 1.2$ & $5.8 \pm 0.6$ & (1) \\
\WASP{} & $219.9^{+2.2}_{-2.1}$ & $9.1^{+1.8}_{-1.3}$ & (2) \\
\COROT{} & $568^{+23}_{-22}$ & $40^{+14}_{-10}$ & (2) \\
\enddata
\tablerefs{$^{(1)}$\cite{Butler2017}
$^{(2)}$\cite{Bonomo2017}}
\label{RV jitter table}
\end{deluxetable}

\begin{figure}
\epsscale{1.25}
\plotone{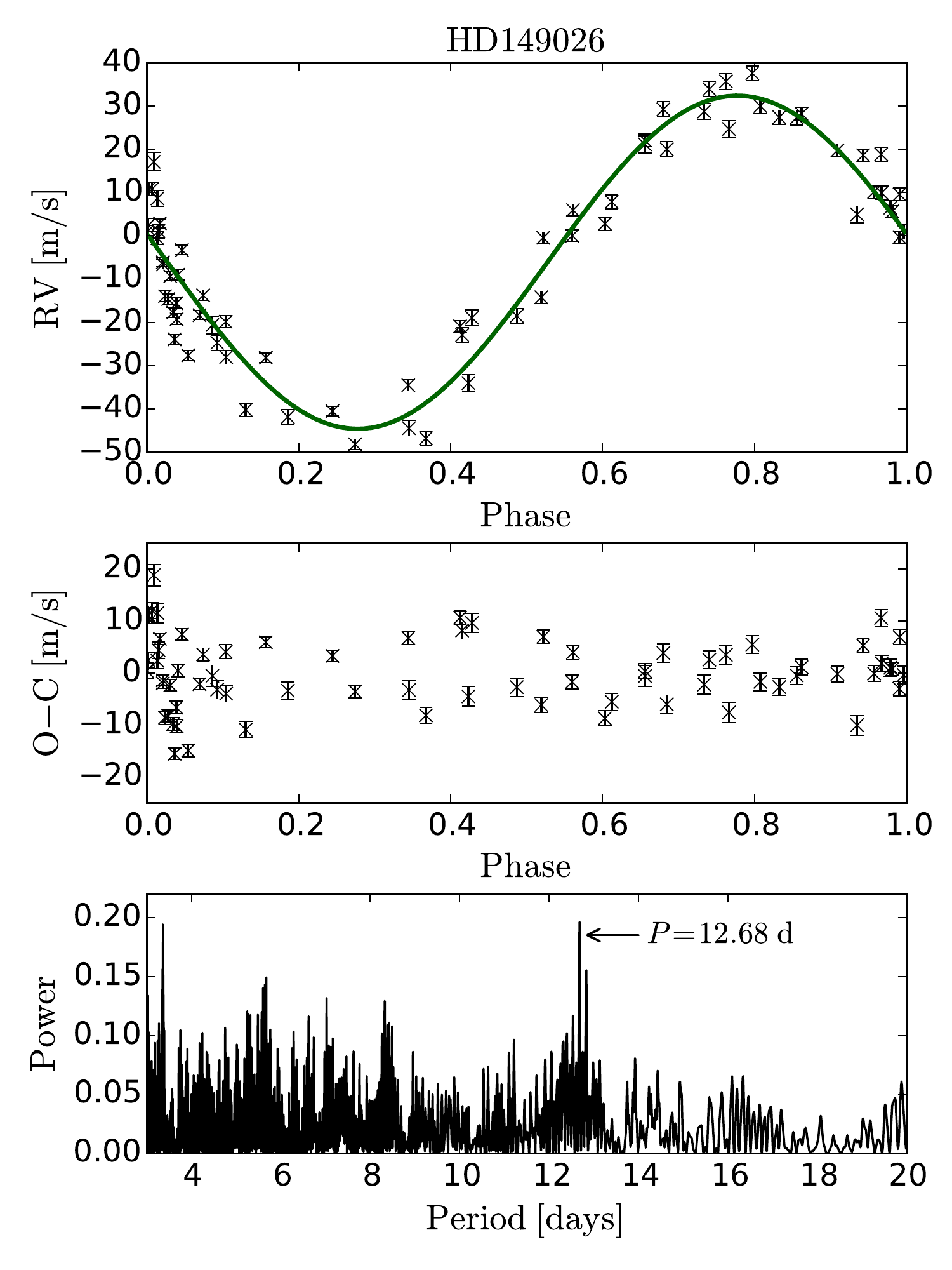}
\caption{\textit{Top panel}: Keck/HIRES Doppler velocity measurements of HD 149026 phase-folded at the period of planet $b$, $P = 2.8759$ days. The solid green line is a circular orbit fit. \textit{Middle panel}: Residuals of the circular orbit fit as a function of orbital phase. \textit{Bottom panel}: Lomb-Scargle periodogram of the residuals, with the highest peak labeled.} 
\label{RV fit, residuals, and periodogram}
\end{figure}

Given its small $\sigma_{\mathrm{jit}}$, it is worthwhile to examine the RVs of HD 149026 to search for any hints of a second signal. The Lick-Carnegie Exoplanet Survey Team (LCES) collected 70 measurements of Keck/HIRES Doppler velocities \citep{Butler2017} over 8.5 years. The top panel of Figure \ref{RV fit, residuals, and periodogram} shows the Keck RVs phase-folded at the period of planet $b$, $P = 2.8759$ days. The second panel shows the residuals obtained after fitting a circular orbit. Finally, the bottom panel displays a Lomb-Scargle periodogram of the RV residuals. There is a noticeable peak at 12.68 days. This peak may be related to stellar activity, since it is close to the $\sim\!13$ day stellar rotation period suggested by the $v \sin i$ measurement from \cite{Sato2005}. Alternatively, it may be the signature of an additional planet, and a closer examination of both the rotational signature in the existing spectra, as well as additional Doppler measurements of the star may be warranted.

\subsection{Summary of Constraints}
In summary, the combination of the requirements of a secular spin-orbit resonance for planet b, total angular momentum conservation, and RV detection limits place strong constraints on the parameters of hypothetical perturbing planets. The limits from angular momentum conservation are particularly restrictive, though we used very conservative assumptions in those calculations. All three systems could therefore host additional, as-yet-undetectable planets with parameters appropriate for generating high-obliquity Cassini states for their companion giant planets. 

\section{Discussion}\label{sec:discussion}
While most thermal full-phase light curves of close-in giant planets show that the hottest region on the planet is eastward of the sub-stellar point, which is consistent with super-rotating winds, there are now at least two planets (\HD{} and \COROT{}) with significant westward hotspot offsets in at least one \textit{Spitzer} band \citep{zha18,Dang2018}. One way of reconciling these westward offsets is to consider variations in the spin axis orientation, which fundamentally changes the relationship between the instellation variations due to the rotation and those due to the orbit. Planets on very short-period orbits occupy an interesting regime where we expect the rotation and orbital rates to be comparable; changes in the spin geometry can therefore have major effects on the observed phases. In this work, we developed a thermal radiative model to investigate how non-zero planetary obliquities may produce observable signatures in full-orbit phase curves.  

It is important to note that we have not considered the degeneracies inherent to a modeling approach that considers only the output 1-D photometry. Many advances have been made recently that focus on first constructing physical 2- or 3-dimensional maps of the planetary flux, then calculating photometric variations \citep[e.g.~][]{Rauscher2018}. These will likely be the most rigorous frameworks for future photometric modeling. Despite these limitations, and acknowledging that there are multiple potential explanations for westward offsets, we showed that, to first order, highly-oblique hot Jupiters should have different phase variation morphologies from those with little to no obliquity.

We focused on a case study of three planets --- \HD{}, \WASP{}, and \COROT{} --- all of which have previous light-curve observations that exhibit anomalous features \citep{cow12,zha18,Dang2018,Barstow2018} that we propose may arise from non-zero obliquities. We reanalyzed the existing \textit{Spitzer} photometry using a thermal model that allows the spin orientation to vary freely. 

A combination of sufficiently long thermal timescales and spin axes nearly perpendicular to the orbit normal can reproduce the strong westward offset in the 3.6 $\mu$m data of \HD{}. However, these oblique models are not preferred over a non-oblique model with fewer parameters, where the westward offset comes from an effective rotation slower than synchronous. The results are quite similar for \COROT{}, where offset-capable models are strongly preferred, and oblique models can effectively reproduce the offset with a very tight constraint on the axial orientation, but due to the slight increase in the number of parameters are statistically not preferred over the simpler non-oblique model. For \WASP{}, variations near quadrature cannot be effectively reproduced by a simple thermal model even with free spin geometry, though they fare better than models with constrained rotation. This suggests that more complex models with tidal distortion or refined instrumental characterization need to be considered to help explain the phase curves of \WASP{}.

Among the three planets we studied, \COROT{} has the tightest constraints on its proposed obliquity, at $45.8^{\circ} \pm 1.4^{\circ}$. In addition to the westward hotspot offset \citep{Dang2018}, a high obliquity state could also account for the planet's extremely inflated radius, which is more anomalous than typical hot Jupiters \citep{Guillot2011}. The enhanced tidal dissipation that accompanies a high obliquity state is more than enough to provide the heat. To second order in eccentricity, the rate of tidal dissipation in a state of equilibrium rotation and according to equilibrium tide theory is \citep{Levrard2007}
\begin{equation}
\frac{\dot{E}_{\mathrm{tide}}(e,\epsilon)}{K} = \frac{2}{1+\cos^2\epsilon}[\sin^2\epsilon + e^2(7+16\sin^2\epsilon)].
\label{dissipation rate}
\end{equation}
Here $K$ is given by 
\begin{equation}
K = \frac{3n}{2}\frac{k_2}{Q_n}\left(\frac{G {M_{\star}}^2}{R_{\rm p}}\right)\left(\frac{R_{\rm p}}{a}\right)^6.
\label{tidalK}
\end{equation}
Using $k_2 = 0.1$ and $C = 0.2$ as in section \ref{sec:Cassini state}, $Q = 5\times10^5$, $e=0$, and $\epsilon = 45.8^{\circ}$ as suggested by the sub-synchronous fit, the tidal dissipation rate is $\dot{E}_{\mathrm{tide}} = 1.2 \times 10^{28} \ \mathrm{erg \ s^{-1}}$. This rate is a substantial fraction of the $2.7 \times 10^{29} \ \mathrm{erg \ s^{-1}}$ of energy from incident stellar radiation, which indicates that obliquity tides may provide a clean solution to the anomalous radius inflation. Although the thermal phase curve modeling of \WASP{} was inconclusive due to tidal distortion complications, this planet also has a highly inflated radius, which might similarly be a product of obliquity tides \citep{MillhollandLaughlin2018}. Variations in the measured optical eclipse depths for \WASP{} point to a possible additional occultation from escaping material \citep{VonEssen2019}, which could correspondingly cause unexpected thermal variability.

If any of these planets do indeed have non-zero obliquities, these configurations could be induced by secular spin-orbit resonances with as-yet undetected exterior planets. We outlined dynamical arguments for each system to identify the regions of parameter space where such obliquity-producing perturbing planets could exist. There is allowable parameter space in each system, particularly in the HD 149026 system. Such results leave open the possibility that these close-in planets may be locked in a stable, high-obliquity spin states.

\acknowledgements
S.M. is supported by the National Science Foundation Graduate Research Fellowship Program under Grant Number DGE-1122492. This material is based upon work supported by the National Aeronautics and Space Administration through the NASA Astrobiology Institute under Cooperative Agreement Notice NNH13ZDA017C issued through the Science Mission Directorate. We acknowledge support from the NASA Astrobiology Institute through a cooperative agreement between NASA Ames Research Center and Yale University.

This research has made use of the NASA Exoplanet Archive, which is operated by the California Institute of Technology, under contract with the National Aeronautics and Space Administration under the Exoplanet Exploration Program.

\facility{Exoplanet Archive, Kepler, Spitzer (IRAC)}

\software{Astropy \citep{ast13}, Colorpy \citep{Kness2008}, python-colormath \citep{Taylor2018}, Jupyter \citep{klu16}, Matplotlib \citep{hun07}, Numpy \citep{van11}, Paletton \citep{pal18}, REBOUND \citep{rei12,rei15}, Scipy \citep{jon01}, WebPlotDigitizer \citep{ank17}}

\bibliographystyle{aasjournal}
\bibliography{apj-jour,paper}

\end{document}